\documentclass[a4paper,12pt]{article}
\usepackage{epsfig}
\usepackage{graphicx}
\usepackage{dcolumn}
\usepackage{bm}
\usepackage{longtable}
\usepackage{amssymb}
\usepackage{amsfonts}
\usepackage[margin=1.0in]{geometry}
\usepackage{amsmath}


\usepackage{slashed}

\def\e{\epsilon}

\DeclareFontFamily{U}{rsf}{}
\DeclareFontShape{U}{rsf}{m}{n}{
  <5> <6> rsfs5 <7> <8> <9> rsfs7 <10-> rsfs10}{}
\DeclareMathAlphabet\Scr{U}{rsf}{m}{n}

\makeatletter                             
\@addtoreset{equation}{section}                   
\makeatother                              
\usepackage{hyperref}

\begin{document}
\begin{center}
{\bf\LARGE On Extremal Limits and Duality Orbits of Stationary Black Holes} \\
\vskip 2 cm
{\bf  Laura Andrianopoli,  Antonio Gallerati and Mario Trigiante}
\vskip 8mm
 \end{center}
\noindent {\small{\it DISAT, Politecnico di Torino, Corso Duca
    degli Abruzzi 24, I-10129 Turin, Italy and Istituto Nazionale di
    Fisica Nucleare (INFN) Sezione di Torino, Italy}

\begin{abstract}
With reference to  the effective three-dimensional description of stationary, single center solutions to (ungauged) symmetric supergravities, we complete a previous analysis  on the definition of a general geometrical mechanism
for connecting global symmetry orbits (duality orbits) of non-extremal solutions to those of  extremal black holes. We focus our attention on a generic representative of these orbits, providing its explicit description in terms of $D=4$ fields.
 As a byproduct, using a new characterization of the angular momentum in terms of quantities intrinsic to the geometry of the $D=3$ effective model, we are able to prove on general grounds its invariance, as a function of the boundary data, under the $D=4$ global symmetry. In the extremal under-rotating limit it becomes moduli-independent.  We also discuss the issue of the fifth parameter characterizing the four-dimensional seed solution, showing that it can be generated by a transformation in the global symmetry group which is manifest in the $D=3$ effective description.
\end{abstract}


\section{Introduction}
\label{Introduction}
The seminal work by \cite{Breitenlohner:1987dg} has defined an effective $D=3$ description
 of (asymptotically flat) stationary black holes in $D=4$ supergravity theories \cite{reviews}, which
 unveiled a larger global symmetry (to be dubbed \emph{duality} in the following) underlying
 these solutions. In fact this approach has provided a valuable tool for their classification
 \cite{Cvetic:1995kv,pioline,Gaiotto:2007ag,Bergshoeff:2008be,Bossard:2009at,Kim:2010bf,Fre:2011uy,bossard2,Chemissany:2012nb}
and consists in describing this kind of solutions as solutions to an effective $D=3$
Euclidean sigma-model which is formally obtained by reducing the $D=4$ theory along the
time direction and dualizing the vector fields into scalars. The action of the global
symmetry group (\emph{duality group}) $G$ of this Euclidean model has been extensively used in the literature as a \emph{solution-generating technique} to construct  non-extremal, rotating, electrically charged black hole solutions coupled to scalar fields
 \cite{Cvetic:1995kv} and, more recently,
  found application in the context of \emph{subtracted geometry} \cite{Bertini:2011ga,Cvetic:2011dn,Virmani:2012kw,Cvetic:2013cja}.\par
Stationary, asymptotically flat, black holes can therefore be conveniently classified
 in orbits with respect to the action of $G$. We shall restrict ourselves here to the
 single-center case. In a recent paper \cite{Andrianopoli:2013kya} we defined  a general geometrical mechanism for connecting the orbit corresponding to non-extremal solutions to those defining the extremal (i.e. zero-temperature) ones, and applied it, as a worked-out example, to the $T^3$-model. Here we wish to complete this analysis by applying the same mechanism to explicit solutions to the STU model, thus proving it for the broad class of symmetric extended supergravities which share the STU model as a common universal truncation. These include all the extended (i.e. $\mathcal{N}\ge 2$) four-dimensional models whose scalar manifold is symmetric of the form $G_4/H_4$, and the isometry group $G_4\subset G$, which defines the global symmetry (or $D=4$-duality) of the four-dimensional theory, is a \emph{non-degenerate group of type-${\rm E}_7$} \cite{brown}.\footnote{In the $\mathcal{N}=2$ case, the above condition in referred to the special K\"ahler manifold spanned by the scalar fields in the vector multiplets, since those in the hypermultiplets are not relevant to the black hole solutions under consideration. Moreover by specializing to the \emph{non-degenerate} case (see the second of references \cite{brown}), we are excluding those models with $G_4={\rm U}(p,q)$ and vector  field-strengths together with their magnetic duals  transforming in the ${\bf p+q}+\overline{\bf p+q}$, like the \emph{minimal coupling}  $\mathcal{N}=2$ models  with $G_4={\rm U}(1,q)$ or the $\mathcal{N}=3$ supergravity  with $G_4={\rm U}(3,q)$. } Those models typically have a $D=5$ uplift and include the maximal and half-maximal  supergravity ($\mathcal{N}=8,\, 4$), the so called ``magical'' $\mathcal{N}=2$ supergravities and the infinite series of models with special K\"ahler manifold $\frac{{\rm SL}(2,\mathbb{R})}{{\rm SO}(2)}\times \frac{{\rm SO}(2,n)}{{\rm SO}(2)\times{\rm SO}(n)}$. At least as far as the single-center solutions are concerned, the $G$-orbits of regular black holes in all these models have a representative in the STU truncation.\par
General features of a stationary  solution, like its rotation and extremality (related
 to the temperature), are in particular associated with invariants of $G$. As far as
 the rotational property of the black hole is concerned, this statement was proven in
 \cite{Andrianopoli:2012ee} by defining a matrix $Q_\psi$ which, just like the Noether charge matrix $Q$, lies
 in the Lie algebra $\mathfrak{g}$ of $G$, and which vanishes if and only if the solution is static. In terms
 of $Q$ and $Q_\psi$ the regularity condition for the black hole solution was written in a $G$-invariant way.
 The matrix  $Q_\psi$ allows to easily infer how the angular momentum $M_\varphi$ transforms under $G$, without
 having to derive the full transformed solution and compute the Komar integral on it. These tools were then applied
 in \cite{Andrianopoli:2013kya} in order to define the general algebraic procedure for connecting the orbit
 of non-extremal solutions to those of extremal ones. In particular,
  as far as extremal under-rotating and static black holes are concerned, this mechanism makes use of
   \emph{singular Harrison transformations} and generalizes previous results in the literature  \cite{Cvetic:1995kv,Rasheed,Larsen,Astefanesei},
   related to  specific electric-magnetic frames. We shall complete this analysis in the present note, by applying it to explicit representatives of the relevant orbits of the global symmetry group $G$,  solutions to the STU model. \par
As a byproduct, using the general expression of $M_\varphi$ in terms of $Q_\psi$,  we are able to prove on general grounds its invariance under the $D=4$ global symmetry, as a function of the boundary values of the scalar fields and of the electric-magnetic charges. In the extremal under-rotating limit the attractor mechanism seems to involve the angular momentum as well: It becomes moduli independent and thus it is only expressed in terms of the quartic invariant of the $D=4$ duality group.
   \par
In the final stage of preparation of the present paper, we became aware of
\cite{Chow:2013tia} where the general rotating black hole solution to the STU
 model was constructed\footnote{The explicit solutions used in the present paper were derived independently.}. Here however we are interested in defining a general
  mechanism for connecting regular black hole $G$-orbits. Since working with the most general solution would somewhat conceal
such a mechanism, we choose to work with the simplest representatives of these orbits (i.e. the most general solutions modulo $G$-transformations or  \emph{seed} solutions with respect to $G$), the generic solution in each orbit being then obtained from them though the action of $G$.\par
We also show that the fifth parameter characterizing the seed solutions with respect to the $D=4$ global symmetry group can be generated by means of $G$.\par
The paper is organized as follows. \\
In Section \ref{sec1} we set the stage for our analysis by recalling the main facts about the effective $D=3$ description of stationary four-dimensional solutions and we define the matrix $Q_\psi$ associated with the rotation of the black hole.\\
In Sect. \ref{kf} we deal with the $G$-orbits of non-extremal and extremal solutions reviewing the general geometrical procedure for connecting them. We also relate important physical properties of the solutions to $G$-invariant quantities associated with the corresponding orbit and outline general features of the matrix $Q_\psi$ in the Kerr orbit, which were not given in our earlier paper. In particular, in the final paragraph of the Section, we prove, using the general expression of $M_\varphi$ in terms of the matrix $Q_\psi$, that on a generic solution the angular momentum \emph{is a $G_4$ invariant function of the values of the $D=4$ scalar fields at radial infinity and of the electric-magnetic charges.} \\
In Sect.s \ref{p0qi} and \ref{q0pi} we give the explicit form of the non-extremal rotating solutions to the STU model (with fully integrated vector fields)  corresponding  to the sets of charges $p^0,\,q_i$ and $q_0,\,p^i$, $i=1,2,3$, and study their limits to extremal static and under-rotating black holes.
We conclude with a discussion of the $5^{th}$ invariant-parameter of a generic $D=4$ single-center solution, with respect to the $D=4$ global symmetry group,  showing that it is not $G$-invariant and that it can be thus generated by means of a $G$-transformation not belonging to $G_4$ (an explicit calculation is given in Appendix \ref{5par}). Consequently the most general extremal, single-center solution to the $D=3$ effective model, modulo $G$-transformations, is a 4-parameter one.

\section{Stationary Single-center Solutions}\label{sec1}
We shall be working with a $D=4$ extended (i.e. $\mathcal{N}>1$), ungauged supergravity,  whose bosonic sector consists in $n_s$
 scalar fields $\phi^r(x)$, $n_v$ vector fields $A^\Lambda_\mu(x)$, $\Lambda=1,\dots, n_v$, and the graviton $g_{\mu\nu}(x)$, which are described by the following Lagrangian
\footnote{Here we adopt the notations and conventions of \cite{Andrianopoli:2012ee,Andrianopoli:2013kya} (in particular we use the ``mostly plus'' convention and $8\pi G={\bf \mathtt{c}}=\hbar=1$).}:
\begin{equation}
\mathcal{L}_4=e\,\left(\frac{R}{2}-\frac{1}{2}\,G_{rs}(\phi^t)\,\partial_\mu \phi^r\,\partial^\mu \phi^s+\frac{1}{4}\,I_{\Lambda\Sigma}(\phi^r)\,F^\Lambda_{\mu\nu} \,F^{\Sigma\,\mu\nu}+ \frac{1}{8\,e}\,R_{\Lambda\Sigma}(\phi^r)\,\epsilon^{\mu\nu\rho\sigma}\,F^\Lambda_{\mu\nu} \,F^{\Sigma}_{\rho\sigma}\right)\,,
\end{equation}
where $e:=\sqrt{|{\rm det}(g_{\mu\nu})|}$. In symmetric supergravities, which we shall restrict to,
 the scalar fields $\phi^s$ span a homogeneous, symmetric, Riemannian scalar manifold:
\begin{equation}
\mathcal{M}^{(4)}_{scal}=\frac{G_4}{H_4}\,,\label{M4}
\end{equation}
where the isometry group  $G_4$ is the symmetry group of the whole theory provided its non-linear action on the
scalar fields is combined with a symplectic action, defining a representation ${\bf R}$ of $G$, on the vector
 field strengths $F^\Lambda=dA^\Lambda$ and their magnetic duals $G_\Lambda$.\par
We shall be dealing with stationary, axisymmetric, asymptotically flat, single center solutions whose space-time metric, in a suitable system of coordinates, has the general form:
 \begin{equation}
ds^2=-e^{2U}\,(dt+\omega_\varphi\,d\varphi)^2+e^{-2U}\,g_{ij}\,dx^i\,dx^j\,,\label{ds2}
\end{equation}
where $i,j=1,2,3$ label the spatial coordinates $x^i=(r,\theta,\varphi)$ and $U,\, \omega_\varphi,\, g_{ij}$ are all
functions of $r,\,\theta$.
The two Killing vectors are $\xi=\frac{\partial}{\partial t}$ and $\psi=\frac{\partial}{\partial \varphi}$.\par
As mentioned in the introduction, these solutions can be given an effective description in an Euclidean $D=3$ model describing gravity coupled to $n=2+ n_s+2n_v$ scalar fields $\phi^I(r,\theta)$ comprising, besides the $D=4$ scalars $\phi^s$, the warp function $U$ and $2n_v+1$ scalars  $\mathcal{Z}^M=\{\mathcal{Z}^\Lambda,\,\mathcal{Z}_\Lambda\}$ and $a$ originating from the time-like dimensional reduction of the $D=4$ vectors and the dualization of the Kaluza-Klein vector $\omega_\varphi$ into a scalar.
The precise relation between the scalars $a,\, \mathcal{Z}^M$ and the four-dimensional fields is \cite{Andrianopoli:2013kya}:
\begin{eqnarray}
A^\Lambda&=& A^\Lambda_0 (dt + \omega)+ A^\Lambda_{(3)} \,,\quad A^\Lambda_{(3)}\equiv A^\Lambda_i dx^i\,,\label{AF1}\\
\mathbb{F}^M&=&\left(
                       \begin{array}{c}
                         F^\Lambda_{\mu\nu} \\
                         G_{\Lambda\,\mu\nu} \\
                       \end{array}
                     \right)\,\frac{dx^\mu\wedge dx^\nu}{2}
                     =d \mathcal{Z}^M \wedge (dt + \omega) + e^{-2 U}\mathbb{C}^{MN}\mathcal{M}_{(4) NP} {}^{*_3} d\mathcal{Z}^P\,,\label{AF2}\\
da&=& - e^{4 U} \,{}^{*_3}d\omega - \mathcal{Z}^T \mathbb{C} d\mathcal{Z}\,,\label{AF3}
\end{eqnarray}
where $*_3$ is the Hodge operation in the $D=3$ Euclidean space, $\mathcal{M}_{(4)}$ the symmetric, symplectic matrix characterizing the symplectic structure over $\mathcal{M}^{(4)}_{scal}$ (see Appendix \ref{stumodel} for an explicit construction).
The effective $D=3$ Lagrangian describes a sigma-model coupled to gravity and reads:
\begin{align} \label{geodaction}
\frac{1}{e^{(3)}}\,\mathcal{L}_{3} &= \frac{1}{2}\,\mathcal{R} - \tfrac{1}{2}
G_{ab}(z)\partial_i{z}^a\partial^i{z}^b=\nonumber\\
&=\frac{1}{2}\, \mathcal{R} -[ \partial_i U \partial^i
U+\tfrac{1}{2}\,G_{rs}\,\partial_i{\phi}^r\,\partial^i{\phi}^s +
\tfrac{1}{2}\e^{-2\,U}\,\partial_i{\mathcal{Z}}^T\,\mathcal{M}_{(4)}\,\partial^i{\mathcal{Z}} +\nonumber\\ &+
\tfrac{1}{4}\e^{-4\,U}\,(\partial_i{a}+\mathcal{Z}^T\mathbb{C}\partial_i{\mathcal{Z}})(\partial^i{a}+\mathcal{Z}^T
\mathbb{C}\partial^i{\mathcal{Z}})]\,,
\end{align}
where $e^{(3)}\equiv \sqrt{{\rm det}(g_{ij})}$ and $\mathbb{C}$ is the symplectic-invariant, antisymmetric matrix.
The scalar fields span a homogeneous, symmetric, pseudo-Riemannian manifold of the form
\begin{equation}
\mathcal{M}_{scal}=\frac{G}{H^*}\,,
\end{equation}
containing $\mathcal{M}^{(4)}_{scal}$ as a submanifold. The isometry group $G$ is a semisimple, non-compact Lie group
which defines the global symmetry of the model, while $H^*$ is a non-compact real form of the maximal compact subgroup of $G$.\par
Stationary axisymmetric solutions are described by $n$ functions $\phi^I(r,\theta)$, solutions to the sigma model equations, and characterized by a unique ``initial point'' $\phi_0\equiv (\phi_0^I)$ at radial infinity
\begin{equation}
\phi_0^I=\lim_{r\rightarrow \infty} \phi^I(r,\theta)\,,
\end{equation}
and an ``initial velocity'' $Q$, at radial infinity, in the tangent space $T_{\phi_0}[\mathcal{M}_{scal}]$, which is the Noether charge matrix of the solution.  Since the action of $G/H$ on $\phi_0$ is transitive, we can always fix $\phi_0$ to coincide with the origin $O$ (defined by the vanishing values of all the scalars) and then classify the orbits of the solutions under the action of $G$ (i.e. in maximal sets of solutions connected through the action of $G$) in terms of the orbits of the velocity vector $Q\in T_{O}(\mathcal{M}_{scal})$ under the action of $H^*$.
The Noether charge matrix $Q$ is computed as:
\begin{equation}
Q=\frac{1}{4\pi}\,\int_{S_2}{}^{*_3}J\,,\label{nc}
\end{equation}
$J=J_i\,dx^i$ being the Noether current. The explicit form of $J$ is given by the standard theory of sigma models on coset manifolds:\begin{equation}
J_i\equiv \frac{1}{2}\partial_i \phi^I\,\mathcal{M}^{-1}\partial_I
\mathcal{M}\,,\label{curr}
\end{equation}
where $\mathcal{M}(\phi^I)=\mathbb{L}(\phi^I)\eta \mathbb{L}(\phi^I)^\dagger$ is an $H^*$-invariant symmetric matrix built out of  the representative  $\mathbb{L}(\phi^I)$ of $G/H$ at the point $\phi^I$ and $\eta$ is a suitable $H^*$-invariant matrix in the chosen representation of $G$ (see Appendix A of \cite{Andrianopoli:2013kya} for the definition of the adopted conventions).\footnote{The coset geometry is defined by  the
 involutive automorphism $\sigma$ on the algebra
$\mathfrak{g}$ of $G$ which leaves the algebra $\mathfrak{H}^*$
generating $H^*$ invariant. All the formulas related to the group $G$ and its generators are referred to a matrix representation of $G$ (we shall in particular use the \emph{fundamental one}). The involution $\sigma$ in the chosen representation has the general action: $\sigma(M)=-\eta M^\dagger\eta$, $\eta$
being an $H^*$-invariant metric ($\eta=\eta^\dagger,\,\,\eta^2={\bf 1}$), and induces the (pseudo)-Cartan
decomposition of $\mathfrak{g}$ of the form:
\begin{equation}
\mathfrak{g}=\mathfrak{H}^*\oplus \mathfrak{K}^*\,,\label{pseudoC}
\end{equation}
where $\sigma(\mathfrak{K}^*)=-\mathfrak{K}^*$, and the following
relations hold
\begin{equation}[\mathfrak{H}^*,\mathfrak{H}^*]\subset\mathfrak{H}^*,
\quad [\mathfrak{H}^*,\mathfrak{K}^*]\subset \mathfrak{K}^*,\quad
[\mathfrak{K}^*,\mathfrak{K}^*] \subset
\mathfrak{H}^*.\label{HKrels}\end{equation}}
The scalar fields $\phi^I$ define a local \emph{solvable parametrization} of the coset, and the coset representative  is chosen to be
\begin{equation}
\mathbb{L}(\phi^I)=\exp(-a T_\bullet)\,\exp(\sqrt{2}
\mathcal{Z}^M\,T_M)\,\exp(\phi^r\,T_r)\,\exp(2\,U
H_0)\,,\label{cosetr3}
\end{equation}
where $T_{\mathcal{A}}=\{H_0,\,T_\bullet,\,T_s,\,T_M\}$ are the solvable generators defined in  Appendix A of \cite{Andrianopoli:2013kya}.\footnote{The structure of this solvable algebra is the following:
\begin{align}
[H_0,\,T_M]&=\frac{1}{2}\,T_M\,\,;\,\,\,[H_0,\,T_\bullet]=T_\bullet\,\,;\,\,\,[T_M\,T_N]=\mathbb{C}_{MN}\,T_\bullet\,,\nonumber\\
[H_0,T_r]&=[T_\bullet,T_r]=0\,\,;\,\,\,[T_r,T_M]=T_r{}^N{}_M\,T_N\,\,;\,\,\,[T_r,T_s]=-
T_{rs}{}^{s'} T_{s'}\,,\label{relc1}
\end{align}
$T_r{}^N{}_M$ representing the symplectic ${\bf R}$ representation of $T_r$ on
contravariant symplectic vectors $d\mathcal{Z}^M$.}
Since the generators $T_M$ transform under the adjoint action of $G_4\subset G$ in the symplectic duality representation ${\bf R}$ of the electric-magnetic charges, we shall use for them the following notation: $(T_M)=(T_{q_\Lambda},\,T_{p^\Lambda})$.\par
The Noether matrix $Q$ encodes all the conserved physical quantities associated with the solution, except the angular momentum $M_\varphi$. In other words it contains no information about the rotation of the solution. In \cite{Andrianopoli:2012ee} we defined a new matrix $Q_\psi$ which describes the global rotation of the solution:
\begin{equation}
Q_\psi =-\frac{3}{4\pi}\,\int_{S_2^\infty}\psi_{[i}\,J_{
j]}\,dx^i\wedge dx^j=\frac{3}{8\pi}\,\int_{S_2^\infty} g_{\varphi\varphi}\,J_\theta\,d\theta d\varphi\,.\label{qupsi}
\end{equation}
The ADM-mass $M_{ADM}$, NUT-charge $n_{NUT}$, electric and magnetic charges $\Gamma^M=(p^\Lambda,\,q_\Lambda)$, scalar charges $\Sigma_s$ and angular momentum $M_\varphi$, associated with the solution are then obtained as components of $Q$ and $Q_\psi$ \cite{Chemissany:2010zp,Andrianopoli:2012ee,Andrianopoli:2013kya}:\footnote{Eq.s (\ref{allthat}) hold also for generic values of the scalar fields at radial infinity, i.e. for $Q,\,Q_\psi\,\in \, T_{\phi_0}[\mathcal{M}_{scal}]$.}
\begin{align}
M_{ADM}&=k\,{\rm
Tr}(H_0^\dagger\,Q)\,,\,\,\,n_{NUT}=-k\,{\rm
Tr}(T_\bullet^\dagger\,Q)\,,\,\,\,\Gamma^M=\sqrt{2}\,k\,\mathbb{C}^{MN}\,{\rm
Tr}(T_N^\dagger\,Q)\,,\,\,\,\Sigma_s=k\,{\rm
Tr}(T_s^\dagger\,Q)\nonumber\\
M_\varphi&=k\,{\rm Tr}(T_\bullet^\dagger\,Q_\psi)\,,\label{allthat}
\end{align}
$k=1/(2{\rm Tr}(H_0H_0))$ being a representation-dependent constant.
Both $Q$ and $Q_\psi$ are matrices in the Lie algebra $\mathfrak{g}$ of $G$. More specifically they belong to the space $\mathfrak{K}^*$ complement in $\mathfrak{g}$ to the algebra $\mathfrak{H}^*$ of $H^*$ and isomorphic to $T_{O}(\mathcal{M}_{scal})$.
\par
Being $G$ the global symmetry group of the effective model, a generic element $g$ of it maps a solution $\phi^I(r,\theta)$ into an other solution $\phi^{\prime\,I}(r,\theta)$ according to the matrix equation:
\begin{equation}
\mathcal{M}(\phi^{\prime I}(x^i))=g\,\mathcal{M}(\phi^{I}(x^i))\,g^\dagger\,.\label{sgt}
\end{equation}
From their definitions (\ref{nc}), (\ref{qupsi}), and from (\ref{sgt}), it follows that $Q$ and $Q_\psi$ transform under the
 adjoint action of $G$ as:
\begin{equation}
\forall g\in G\,\,:\,\,\,\,Q\rightarrow Q'=(g^{-1})^\dagger\,Q\,g^{\dagger}\,\,\,;\,\,\,\,\,Q_\psi\rightarrow
 Q'_\psi=(g^{-1})^\dagger\,Q_\psi\,g^{\dagger}\,.\label{QQpsitra}
\end{equation}
Eq.s (\ref{allthat}), and the last one
 in particular, allow to compute the angular momentum of the transformed solution without having to explicitly derive the
 latter from (\ref{sgt}) and to compute the corresponding Komar integral on it. This is one of the main advantages of working with $Q_\psi$.
The presence of a non-vanishing $Q_\psi$ is a characteristic of the $G$-orbits
of rotating solutions and therefore one cannot \emph{generate} rotation on a static  $D=4$ solution using $G$ !
\section{The Kerr Family}\label{kf}
As proven in \cite{Breitenlohner:1987dg},
the most general (non-extremal) stationary,  axisymmetric  single black hole solution to
the model can be obtained from the Kerr solution through a $G$-transformation (more precisely through a Harrison transformation).
The matrices $Q$ and $Q_\psi$ for the Kerr solution, characterized by a mass $m$ and an angular-momentum parameter $\alpha$ are
 diagonalizable and thus their $G$-orbits are uniquely characterized by their eigenvalues. In the pure Kerr solution, $Q,\,Q_\psi$ belong,
  modulo multiplication by $\alpha$, to the same $G$-orbit. In fact we have:
\begin{equation}
Q_\psi=\alpha\,h^{-1}\,Q\,h\,\,\,;\,\,\,\,h\in {\rm U}(1)_E\,,\label{QQpsi}
\end{equation}
where ${\rm U}(1)_E$ is the compact Ehlers transformation group.
As we shall see, in the extremal limits we are going to consider, this will no longer be the case.
The matrix  $Q$ belongs to the \emph{Schwarzschild orbit} characterized,
 choosing for $Q$ the fundamental representation of $G$ and if $G\neq {\rm E}_{8(8)},\,{\rm E}_{8(-24)}$,
by the matrix equation \cite{Bossard:2009at}
\footnote{If $G$ is a real form of ${\rm E}_{8}^{\mathbb{C}}$, the fundamental and the adjoint representation coincide and the matrix
equation becomes quintic in $Q$,\cite{Bossard:2009at}.}:\footnote{The constant $\bar{c}^2$ in the case of the Kerr-Newmann-NUT black hole with electric and magnetic
charges $q,\,p$ and NUT charge $n_{NUT}$, reads: $\bar{c}^2=\frac{k}{2}{\rm Tr}(Q^2)=m^2+n_{NUT}^2-\frac{p^2+q^2}{2}$.}
\begin{equation}
Q^3=\bar{c}^2\,Q\,\,\,,\,\,\,\,\bar{c}^2=\frac{k}{2}{\rm Tr}(Q^2)=m^2\,.\label{eqa1}
\end{equation}
From (\ref{QQpsi}) it follows that:
\begin{equation}
Q_\psi^3=\alpha^2\bar{c}^2\,Q_\psi\,\,\,,\,\,\,\,\alpha^2=\frac{{\rm Tr}(Q_\psi^2)}{{\rm Tr}(Q^2)}\,.\label{eqa2}
\end{equation}
Also  the following matrix equations are satisfied:
\begin{equation}
Q_\psi^2\,Q=\alpha^2\bar{c}^2\,Q\,\,;\,\,\,Q^2\,Q_\psi=\bar{c}^2\,Q_\psi\,.\label{eqa3}
\end{equation}
It is worth emphasizing that the equations (\ref{eqa1}), (\ref{eqa2}), , (\ref{eqa3}) together with the trace expression for $m$ and $\alpha$,
are $G$-invariant and thus hold for any representative of the Kerr $G$-orbit.
We can then define an extremality parameter $c^2$ in terms of the following $G$-invariant quantity \cite{Andrianopoli:2012ee}:
\begin{equation}
c^2=m^2-\alpha^2=\frac{k}{2}{\rm Tr}(Q^2)-\frac{{\rm Tr}(Q_\psi^2)}{{\rm Tr}(Q^2)}\,.\label{eqa4}
\end{equation}
In terms of $c$ we can write the \emph{Hawking temperature} of the black hole in the form:
\begin{equation}
T=\frac{1}{2\pi}\frac{c}{\alpha\,|\omega_H|}=\frac{c}{2\,S}\,,\label{temperature}
\end{equation}
where $S$ is the Bekenstein-Hawking entropy   of the solution, expressed in the chosen units, in terms of the horizon
area $A$, by the renown formula
\begin{equation}
S=\frac{k_B \,{\bf \mathtt{c}}^3}{G\hbar}\,\frac{A}{4}=\frac{A}{4}=\pi\,\alpha\,|\omega_H|\,,\label{entropy}
\end{equation}
while $\omega_H$ is defined as:
\begin{equation}
\omega_H:=\lim_{r\rightarrow r_+}\omega_\varphi\,\,\,;\,\,\,\,r_+=m+c\,.
\end{equation}
The above expression allows to write the regularity bound for the Kerr solution \emph{in a $G$-invariant
form} which thus holds for any representative of the Kerr-orbit:
\begin{equation}
m^2\ge \alpha^2\,\,\,\Leftrightarrow\,\,\,\,\frac{k}{2}{\rm Tr}(Q^2)\ge\frac{{\rm Tr}(Q_\psi^2)}{{\rm Tr}(Q^2)}\,.\label{regbound}
\end{equation}
\paragraph{Angular momentum and duality.} Let us comment on the properties of the angular momentum $M_\varphi$ with respect to the four-dimensional duality symmetry $G_4$. In our analysis, for the sake of simplicity, we have fixed the transitive action of $G/H^*$ on the solution by choosing the scalar fields at infinity to correspond to the origin $O$ of the manifold. Let us relax this assumption in the present paragraph. All the formulas given in the previous section, including (\ref{allthat}), clearly hold for generic ``initial values'' of the scalar fields.\par
 In general, on a rotating black hole solution, the angular momentum would depend on the boundary values  $(\phi_0^s)$ of  $(\phi^s)$ and on the electric-magnetic charges $\Gamma^M$ and be expressed in terms of $Q_\psi$ by the last of eq.s (\ref{allthat}). Suppose now we transform the solution by means of an element $g\in G_4$ into another one with boundary values $\phi_0^{\prime s}$ and charges $\Gamma^{\prime M}$
\begin{equation}
g\,\,:\,\,\,\,(\phi_0^s,\,\Gamma^M)\,\,\,\longrightarrow\,\,\,\,\,(\phi_0^{\prime s},\,\Gamma^{\prime M})\,.
\end{equation}
Let us prove, by using the definition in (\ref{allthat}), that $M_\varphi$ is not affected by the action of $g$. The matrix $Q_\psi'$ associated with the new solution is related to $Q_\psi$ by (\ref{QQpsitra}), so that the corresponding angular momentum $M'_\varphi=M_\varphi(\phi_0^{\prime s},\,\Gamma^{\prime M})$ reads:
\begin{equation}
M_\varphi(\phi_0^{\prime s},\,\Gamma^{\prime M})=k\,{\rm Tr}(T_\bullet^\dagger\,Q'_\psi)=k\,{\rm Tr}(T_\bullet^\dagger\,(g^{-1})^\dagger\, Q_\psi\,g^\dagger)=k\,{\rm Tr}(T_\bullet^\dagger\, Q_\psi)=M_\varphi(\phi_0^s,\,\Gamma^M)\,,\label{Minvar}
\end{equation}
where we have used the property that $G_4$ commutes with the Ehlers group  ${\rm SL}(2,\mathbb{R})_E$ inside $G$, so that its elements commute with the $\mathfrak{sl}(2,\mathbb{R})_E$ generators $\{H_0,\,T_\bullet,\,T_\bullet^\dagger\}$.\par
We conclude that $M_\varphi$, \emph{is a $G_4$-invariant function of the scalar fields at radial infinity and the electric-magnetic charges.}
 This is indeed what one would expect for the angular momentum of a solution: being a quantity related to  its
spatial rotation it should not be affected by a $D=4$ duality transformation. \par
Clearly the above derivation would not hold for a generic global symmetry transformation in $G$. As we shall see below, in the under-rotating limit $M_\varphi$ is independent of $\phi_0^s$ and thus is expressed in terms of the $G_4$-invariant of the electric-magnetic charges alone, namely  the quartic invariant function $I_4(p,q)$. A similar thing happens for the horizon area (i.e. the entropy) by virtue of the \emph{attractor mechanism} (see below). We conclude from this observation
 that there seems to be an ``attractor mechanism'' at work also for the angular momentum.\par
 Finally let us notice that the simple proof (\ref{Minvar}) also applies to the ADM-mass and the NUT-charge, both given in (\ref{allthat}).
 This is consistent with the duality invariance  of $M_{ADM}$ proven in \cite{Andrianopoli:2010bj} (see eq. (29) therein) in a different and more sophisticated way.

\section{Extremal Limits}\label{el}
The regularity bound $c^2\ge 0$ is saturated for the  \emph{extremal} solutions, which are thus characterized by a vanishing
 Hawking temperature (\ref{temperature}). This bound can be saturated in essentially two ways:
 \begin{itemize}
 \item Both sides of (\ref{regbound}), though equal, stay different from zero. The extremality condition thus becomes a
 constraint on the two non-vanishing $G$-invariants. The resulting solution is called \emph{over-rotating}
 extremal and retains, in this limit, the presence of an ergosphere. The two matrices $Q$ and $Q_\psi$ are still diagonalizable;
 \item Both sides of (\ref{regbound}) vanish separately. The resulting solution can either be extremal
 \emph{under-rotating} \cite{Rasheed,Larsen,Goldstein:2008fq,Bena1,Bena2}
 or extremal-static and has no ergosphere. In this limit \cite{Andrianopoli:2013kya} both $Q$ and $Q_\psi$ become
 nilpotent, \emph{belonging to different $G$-orbits} (or better $H^*$ orbits on $T_O[\mathcal{M}_{scal}]\sim \mathfrak{K}^*$).
 \end{itemize}
 We shall focus on the second limit, which has been considered in the literature in specific contexts:
 Heterotic theory \cite{Cvetic:1995kv,Astefanesei}; Kaluza-Klein supergravity \cite{Rasheed,Larsen}. In
 \cite{Andrianopoli:2013kya} we defined a general geometric prescription for connecting the
 non-extremal Kerr-orbit to the extremal static or under-rotating ones, in a way  which is frame-independent
  (i.e. does not depend on the particular string theory and compactification yielding the four-dimensional supergravity).
  This procedure makes use of \emph{singular Harrison transformations} by means of which an In\"on\"u--Wigner contraction on the matrices
    $Q$ and $Q_\psi$ is effected, resulting in the nilpotent matrices $Q^{(0)}$ and $Q^{(0)}_\psi$ associated with  extremal
    static or under-rotating black holes.\par
 Harrison transformations \cite{Breitenlohner:1987dg} are $H^*$-transformations
 which play a special role in the solution generating techniques: They are not present
 among the global symmetries of the  $D=4$ theory and have the distinctive property of
 switching on electric or magnetic charges when acting on neutral solutions (like the
  Kerr or Schwarzshild ones). Their generators  $(\mathbb{J}_M)=(\mathbb{J}_\Lambda,\,\mathbb{J}^\Lambda)$
   in $\mathfrak{H}^*$ are in one-to-one correspondence with the electric and magnetic charges
    $(\Gamma^M)=(p^\Lambda,\,q_\Lambda)$ and are non-compact (i.e. are represented, in a suitable basis,
    by hermitian matrices). The space ${\rm Span}(\mathbb{J}_M)$ generated by $\{\mathbb{J}_M\}$ is the
    coset space of the symmetric manifold $H^*/H_c$, $H_c$ being the maximal compact subgroup of $H^*$, and thus
it is the carrier of a representation of $H_c$ (the same representation in which the charges
$\Gamma^M$ transform with respect to $H_c$). In general this group has the following structure:  $H_c={\rm U}(1)_E\times H_4$.\par
In \cite{Andrianopoli:2013kya} we considered the \emph{maximal abelian subalgebra} (MASA) of  the space
${\rm Span}(\mathbb{J}_M)$. This is a subspace whose generators $\mathbb{J}^{(N)}=\{\mathcal{J}_\ell\}$
are defined by the \emph{normal form} of the electric and magnetic charges, i.e. the minimal subset of charges into
which the charges of the most general solution can be rotated by means of $H_c$. Its dimension $p$ is therefore just the
\emph{rank of the coset} $H/H_c$. In the maximal supergravity, for example,
 $p={\rm rank}\left(\frac{{\rm SO}^*(16)}{{\rm U}(8)}\right)=4$, the same being true for the half-maximal
 theory, $p={\rm rank}\left(\frac{{\rm SO}(6,2)\times {\rm SO}(2,6+n)}{{\rm SO}(2)^2\times {\rm SO}(6)\times {\rm SO}(6+n)}\right)=4$,
  and for the $\mathcal{N}=2$ symmetric models with rank-3 scalar, special K\"ahler manifold in $D=4$ (for this class of theories, $p=$ rank $+1$).
The simplest representative of the latter class of models is the $STU$ one, which is a consistent truncation
of all the others, besides being a truncation of the maximal and half-maximal theories.
Therefore its space $\mathbb{J}^{(N)}$ is contained in the spaces of Harrison generators
of all the above mentioned symmetric models. As a consequence of this, for the sake of
 simplicity, we can restrict ourselves to the simplest $STU$ model since the $G$-orbits
 of non-extremal and extremal regular solutions to the broad class of symmetric models
  mentioned above  have a representative in the common $STU$  truncation. As for the
  restricted number of $\mathcal{N}=2$ symmetric models for which the rank of
  $\mathcal{M}^{(4)}_{scal}$ is less than 3 ($p< 4$), the following discussion  has a straightforward generalization
   (the $T^3$-model  case with $p=2$ was dealt with in detail in \cite{Andrianopoli:2013kya}).
 Depending on the symplectic frame, i.e. on the higher-dimensional origin of the
  four-dimensional theory, this normal form can consist of different kinds of charges. In all cases
  this normal form can be geometrically  characterized as follows. If we express the Harrison generators in the form:
  \begin{equation}
  \mathbb{J}_M=\frac{1}{2}(T_M+(T_M)^\dagger)=\frac{1}{2}( E_{\gamma_M}+( E_{\gamma_M})^\dagger)\,,
  \end{equation}
  where $\gamma_M$ are the $2n_v$ roots of $\mathfrak{g}$ such that $\gamma_M(H_0)=1/2$, the $p$  generators $\mathcal{J}_\ell$
  are defined by a maximal set $\{\gamma_\ell\}$ of mutually orthogonal roots
  among the $\gamma_M$: $\gamma_{\ell_1}\cdot \gamma_{\ell_2}\propto \delta_{\ell_1,\ell_2}$
  \begin{equation}
  \mathcal{J}_\ell=\frac{1}{2}(E_{\gamma_\ell}+( E_{\gamma_\ell})^\dagger)\,.
  \end{equation}
  \paragraph{Symplectic frames and normal forms.} Since the normal form of the electric and magnetic
  charges with respect to the group $H_c$, for all the symmetric models mentioned above, is contained in the STU truncation, let
  us illustrate within the latter, the relevant symplectic frames.
  The STU model is a $\mathcal{N}=2$ supergravity coupled to three vector multiplets whose three complex scalars span a special
   K\"ahler manifold (\ref{M4}), where $G_4={\rm SL}(2,\mathbb{R})^3$ and $H_4={\rm SO}(2)^3$. Upon time-like reduction to $D=3$, the scalar manifold is enlarged to
    $\mathcal{M}_{scal}=G/H={\rm SO}(4,4)/{\rm SO}(2,2)\times {\rm SO}(2,2)$ (see Appendix \ref{stumodel} for notations and technical details about the STU model).\par
    If the STU model originates from Kaluza--Klein reduction from $D=5$, the resulting symplectic frame corresponds to the following ordering of the
    roots $\gamma_M$, $M=1,\dots,8$:
  \begin{align}
  (\Gamma_M)&=(\mathbb{C}_{MN}\Gamma^N)=(q_\Lambda,\,-p^\Lambda)\leftrightarrow\,\,(\gamma_M)\,,\nonumber\\
  (\vec{\gamma}_a)_{a=1,\dots,4}&=\left[\left(\frac{1}{2},-\frac{1}{2},-\frac{1}{2},-\frac{1}{2}\right),
  \left(\frac{1}{2},\frac{1}{2},-\frac{1}{2},-\frac{1}{2}\right),\,\left(\frac{1}{2},-\frac{1}{2},\frac{1}{2},-\frac{1}{2}\right),\,\left(\frac{1}{2},-\frac{1}{2},-\frac{1}{2},\frac{1}{2}\right)\right]\,,\nonumber\\
  (\vec{\gamma}_{a+4})_{a=1,\dots,4}&=\left[\left(\frac{1}{2},\frac{1}{2},\frac{1}{2},\frac{1}{2}\right),
  \left(\frac{1}{2},-\frac{1}{2},\frac{1}{2},\frac{1}{2}\right),\,\left(\frac{1}{2},\frac{1}{2},-\frac{1}{2},\frac{1}{2}\right),\,\left(\frac{1}{2},\frac{1}{2},\frac{1}{2},-\frac{1}{2}\right)\right]\,,\label{gammaord}
  \end{align}
  where we have represented each root $\gamma_M$ by its component vector $\vec{\gamma}_M$ in a Cartan subalgebra of
  $\mathfrak{so}(4,4)$: The first component is the grading $\gamma_M(H_0)$ with respect to the ${\rm O}(1,1)$ generator $H_0$ in the
  Ehlers group ${\rm SL}(2,\mathbb{R})_E$, the other entries are the components $\gamma_M(H_{\alpha_i})/2$, with respect to the Cartan
  generators $H_{\alpha_i}$ of $G_4$.
  We see that there are two maximal sets of $p=4$ mutually orthogonal roots $\{\gamma_{\ell}\}=\{\gamma_1,\,\gamma_6,\,\gamma_7,\,\gamma_8\}$
  and $\{\gamma_{\ell'}\}=\{\gamma_2,\,\gamma_3,\,\gamma_4,\,\gamma_5\}$, corresponding to the normal forms of the charge vector
  with non-vanishing charges $\{q_0,\,p^i\}_{i=1,2,3}$ and $\{p^0,\,q_i\}_{i=1,2,3}$, respectively.\par
  If we embed the STU model in toroidally compactified
   Heterotic theory \cite{Cvetic:1995kv}, one of the ${\rm SL}(2,\mathbb{R})$s in $G_4$ has a non-perturbative (i.e. not block-diagonal)
   duality action in the ${\bf R}={\bf (2,2,2)}$, while the remaining two factors have a block diagonal symplectic representation.
   The corresponding symplectic frame is characterized by the following order of the roots $\gamma_M$:
   \footnote{This ordering is related to the property that,  in this frame, the Cartan generator of the non-perturbative ${\rm SL}(2,\mathbb{R})$
   be degenerate over the electric (and thus also over the magnetic) charges.}
   \begin{equation}
   (\Gamma^{\prime}_M)\leftrightarrow\,\,(\gamma_1,\gamma_6,\gamma_3,\gamma_4,\gamma_5,\gamma_2,\gamma_7,\gamma_8)\,.
   \end{equation}
   The two normal forms of the charge vector, being identified by the same sets of roots $\{\gamma_\ell\}$ and $\{\gamma_{\ell'}\}$,
    now correspond to two electric and two magnetic charges: $\{p^{\prime 2},p^{\prime 3},
    \,q'_0,q'_1\}$ and $\{p^{\prime 0},p^{\prime 1},\,q'_2,q'_3\}$.\par Finally one can consider the frame in which the
    generators of $G_4$
    can be chosen to be represented  symplectic matrices which are either block diagonal or completely block-off-diagonal (i.e. having entries only in the off-diagonal blocks). This is the frame
    originating from direct truncation of the
$\mathcal{N}=8$ theory in which the ${\rm SL}(8,\mathbb{R})$ subgroup of ${\rm E}_{7(7)}$ has a block-diagonal embedding in
${\rm Sp}(56,\mathbb{R})$.  It corresponds to the following
   order of the roots $\gamma_M$:
   \begin{equation}
   (\Gamma^{\prime\prime}_M)\leftrightarrow\,\,(\gamma_5,\gamma_2,\gamma_3,\gamma_4,\gamma_1,\gamma_6,\gamma_7,\gamma_8)\,.
   \end{equation}
 The two normal forms of the charge vector now correspond to either all electric or all magnetic charges: $\{p^{\prime\prime\,\Lambda}\}$
  and $\{q''_\Lambda\}$.
\par
In all these cases, the MASAs of ${\rm Span}(\mathbb{J}_M)$ are always defined by the \emph{same sets} of generators
$\{\mathcal{J}_\ell\}_{\ell=1,6,7,8},\,\{\mathcal{J}_{\ell'}\}_{\ell'=2,3,4,5}$.
We shall use in the following the first symplectic frame.
\paragraph{The procedure.} Let us summarize the procedure defined in \cite{Andrianopoli:2012ee,Andrianopoli:2013kya} in order to connect the Kerr
 orbit to orbits of  extremal under-rotating and static solutions. We transform the Kerr solution by means
 of a Harrison transformation
 generated by the chosen   MASA $\mathbb{J}^{(N)}$ of ${\rm Span}(\mathbb{J}_M)$:
 \begin{equation}
 \mathcal{O}\in \exp\left(\mathbb{J}^{(N)}\right)\,\,;\,\,\,\,\mathcal{O}=\begin{cases}e^{\sum_{\ell}\log(\beta_\ell)\,\mathcal{J}_\ell}
  & \{q_0\,p^i\}\mbox{ - case}\cr e^{\sum_{\ell'}\log(\beta_{\ell'})\,\mathcal{J}_{\ell'}}
  & \{p^0\,q_i\} \mbox{ - case}\end{cases}\,,
 \end{equation}
 where $\ell=1,6,7,8$ and $\ell'=2,3,4,5$.
 The matrices $Q,\,Q_\psi$ transform according to eq. (\ref{QQpsitra}):
 \begin{equation}
 Q\,\,\rightarrow\,\,\,\,Q'=(\mathcal{O}^{-1})^\dagger\,Q\,\mathcal{O}^\dagger \,\,\,;\,\,\,\,Q_\psi\,\,\rightarrow\,\,;
 \,\,Q'_\psi=(\mathcal{O}^{-1})^\dagger\,Q_\psi\,\mathcal{O}^\dagger\,.
 \end{equation}
  Next we perform, in the two cases, the rescaling:
 \begin{equation}
\beta_\ell\rightarrow m^{\sigma_\ell}\beta_\ell\;\;,\,\,\,(\beta_{\ell'}\rightarrow m^{\sigma_{\ell'}}\beta_{\ell'})\;\;\;\;\alpha\rightarrow m\Omega\,,
\end{equation}
where $\sigma_\ell,\,(\sigma_{\ell'})=\;\pm1$. We then send $m$ to zero. This limit corresponds to an In\"on\"u-Wigner
contraction of $Q'$ and $Q'_\psi$  which
become nilpotent matrices $Q^{(0)},\,Q_\psi^{(0)}$ with a different degree of nilpotency, i.e. belonging to different $H^*$-orbits:
$Q^{(0)}$ has degree three while $Q_\psi^{(0)}$ either vanishes or has degree two. This explains why, in the $m\rightarrow 0$ limit, the ratio on the right hand side of eq. (\ref{regbound})
goes to zero: the numerator ${\rm Tr}(Q_\psi^2)$ vanishes faster than the denominator ${\rm Tr}(Q^2)$.
The charge vector $\Gamma^M$ of the resulting solution, in the two cases, has 4 non-vanishing charges
corresponding to the
chosen normal form, i.e. $ \{q_0\,p^i\}$ or $\{p^0\,q_i\}$.  Depending on the choice of the gradings $\sigma_\ell$
(or $\sigma_{\ell'}$) the charge vector $\Gamma^M$ can belong to any of the $G_4$-orbits of regular solutions,
characterized in terms of the
 $G_4$-quartic invariant $I_4(p,q)$ of the representation ${\bf R}$ as follows \cite{Bellucci:2006xz} (see Appendix \ref{stumodel} for the explicit form of $I_4(p,q)$ in the STU model):
\begin{align}
\mbox{BPS}&:\,\,I_4(p,q)>0\,\,\,\,\mbox{$\mathbb{Z}_3$-symmetry on the  $p^i$ and the $q_i$}\,,\nonumber\\
\mbox{non-BPS}_1&:\,\,I_4(p,q)>0\,\,\,\,\mbox{no $\mathbb{Z}_3$-symmetry}\,,\nonumber\\
\mbox{non-BPS}_2&:\,\,I_4(p,q)<0\,.\nonumber
\end{align}
 For those choices of the gradings yielding $I_4>0$ we find both the BPS and a non-BPS solution
 and the resulting angular momentum is zero (extremal-static black hole, $Q^{(0)}_\psi={\bf 0}$). Only in the cases for which $I_4<0$ we find a rotating solution,
 which is the known under-rotating solution of \cite{Rasheed,Larsen,Goldstein:2008fq,Bena1,Bena2}.
 Therefore we find, as a general result, that the extremal solutions obtained in this way have an angular momentum given by
  \begin{equation}
  M_{\varphi}^{(extr)}=\frac{\Omega}{4}\,\sqrt{|I_4(p,q)|}\,(1-\varepsilon)\,,\label{MI4}
  \end{equation}
  where $I_4=\varepsilon\,|I_4|$ (the above  equation was verified on the 5-parameter solution, see Appendix \ref{5par}). This formula makes the invariance of $M_\varphi$ under  $G_4$-transformations, proven for a generic solution at the end of the previous section,  manifest, since both $I_4(p,q)$ and
  $\Omega=M^{(Kerr)}_{\varphi}/m^2$ are  $G_4$-invariants, being the latter related to the original Kerr solution.
  Actually  on our solutions we cannot see the dependence of the various quantities on the scalar fields $\phi_0^I$, and in particular on the four-dimensional ones, at radial infinity, since these were fixed to zero. Having proven, however, in the previous section that $M_\varphi$ is a $G_4$-invariant function of $\phi_0^s$ and $\Gamma^M$, and having proven on our solutions that it is already an invariant function of the electric-magnetic charges alone, we conclude that, for the under-rotating solutions, $M_\varphi$ only depends on $p^\Lambda,\,q_\Lambda$.\par
   Similarly one finds for the entropy, related to the horizon area and expressed in (\ref{entropy}),
  the following form in the limit:\footnote{The expression $S^{(extr)}=\pi\,\sqrt{|I_4|-4 M_\varphi^{(extr)\,2}}$ is known (see for instance \cite{Bena1}), while the
  last expression, which makes the $G_4$-invariance of $S$ manifest by expressing $M_\varphi^{(extr)}$ in terms of the invariants $I_4,\,\Omega$, to our knowledge,
 is not.}
  \begin{equation}
S^{(extr)}=\pi\,\lim_{m\rightarrow 0} \alpha\,|\omega_H|=\pi\,\lim_{m\rightarrow 0} m\,\Omega\,|\omega_H|=\pi\,\sqrt{|I_4|-4 ( M_{\varphi}^{(extr)})^2}=
\pi\,\sqrt{|I_4|}\sqrt{1-\frac{1}{2}\Omega^2(1-\varepsilon)}\,.\label{SI4}
  \end{equation}
The last expression, obtained by using (\ref{MI4}), makes it manifest that $S^{(extr)}$, as well as the whole near horizon geometry, is
$G_4$-invariant as $M^{(extr)}_\varphi$ is. In the rotating extremal case ($\varepsilon=-1$) we further need to impose $\Omega<1$ in order for the solution
to be well behaved.\par
We observe, however, that before the extremal limit $m\rightarrow 0$ is effected, the expression of  $S$
is not $G_4$-invariant. This can be explained by the fact that we generally made the $G_4$ ``gauge'' choice corresponding to fixing the 4D scalar fields at infinity at the origin of the moduli space, thus breaking the manifest $G_4$ invariance to $H_4$. In the extremal under-rotating and static cases
the \emph{attractor mechanism} is at work \cite{attractors1}, as a consequence of which the near horizon geometry
becomes independent of the values of the scalar fields at radial infinity (which we have fixed to the origin) and only depends on
the  quantized charges $p^\Lambda,\,q_\Lambda$. In the non-extremal case, $c^2>0$, this is no longer the case and the near
 horizon geometry, as well as the entropy, depends on the scalar fields at infinity $\phi^s_0$. We can then argue
 that $S=S(p,q,\phi_0^s)$ is still invariant under $G_4$, provided we transform both $\Gamma^M$ and $\phi_0^s$ simultaneously, just as it was proven at the end of last section to happen for the angular momentum.
 In other words, within our choice of scalar boundary conditions, $S$ is  expressed in terms of $H_4$-invariants and, in the extremal limit, such expression should reduce to
the only scalar-independent $H_4$-invariant, namely to (\ref{SI4}).\par
In the following sections, we work out the explicit solutions to the STU model, corresponding to the two normal forms, the complete
description of which (including the integrated $D=4$ vector fields), to our knowledge, were not present in the literature before
\cite{Chow:2013tia} and which were derived by us independently.
 We then apply to them the general extremal limits discussed above, to derive extremal-static and under-rotating solutions.
 For a detailed algebraic description of the limit $Q',\,Q_\psi'\rightarrow
Q^{(0)},\,Q^{(0)}_\psi$ we refer the reader to Sect. 3 of \cite{Andrianopoli:2013kya}.

\section{The $p^0,\,q_i$\;-\;case}\label{p0qi}
Here we give and discuss the non-extremal, rotating axion-dilaton solution generated by applying to the Kerr solution
the Harrison transformation generated by  $\mathcal{J}_{\ell'} =\{\mathcal{J}_2\,,\mathcal{J}_3\,,\mathcal{J}_4\,,\mathcal{J}_5\}$,
 of the form:
\begin{equation}
\mathcal{O}_{(p^0,\,q_i)}=\;e^{\log(\beta_{\ell'})\mathcal{J}_{\ell'}}\;,
\;\;\;\;\;\;\;\;\;{\ell'}=2,3,4,5\;.
\end{equation}
\newline
The $D=3$ scalars $\Phi^I(r,\theta)$ describing the transformed solution are obtained in terms of
$\Phi^I_{(K)}(r,\theta)$ by solving the matrix equation :
\begin{equation}
\mathcal{M}[\Phi^I(r,\theta)]=\mathcal{O}_{(p^0,\,q_i)}\;\mathcal{M}[\Phi^I_{(K)}(r,\theta)]\;\mathcal{O}_{(p^0,\,q_i)}^T\,.\label{matep0qi}
\end{equation}
\newline
It is convenient, in order to write  $\Phi^I(r,\theta)$, to introduce the following combination of the $\beta$-parameters:
\begin{align}
c_{\ell'}=&\;\frac{1+\beta_{\ell'}}{2\,\sqrt{\beta_{\ell'}}}\;,\;\;\;\;\;\;
s_{\ell'}=\frac{-1+\beta_{\ell'}}{2\,\sqrt{\beta_{\ell'}}}\;\;\;\;\;\;\;\;\;\;(\ell'=2,3,4,5)\nonumber\\
P_c=&\;c_2\,c_3\,c_4\,c_5\;,\;\;\;\; P_s=\;s_2\,s_3\,s_4\,s_5\;.
\end{align}
Recall that, for the Kerr solution, the $D=3$ metric $g^{(3)}=(g_{ij})$ in (\ref{ds2}) can be chosen in  the following general form:
\begin{align}
g^{(3)}=(g_{ij})&={\rm diag}\left(\frac{\tilde{\Delta}}{\Delta},\,\tilde{\Delta},\,\Delta\sin^2(\theta)\right)\,,\nonumber\\
\Delta&=\;(r-m)^2-(m^2-\alpha^2)\,\,;\,\,\,\,\,
\tilde{\Delta}=\;\Delta-\alpha^2\sin^2\theta\,.
\end{align}
We also define the quantity
\begin{align}
\rho^4=&\;\left(\alpha^2\cos^2\theta+(r+2m{s_2}^2)(r+2m{s_3}^2)\right)\,
        \left(\alpha^2 \cos^2\theta +(r + 2 m {s_4}^2) (r + 2 m{s_5}^2)\right)\,-\nonumber\\
        &-4\alpha^2 m^2 (c_2 c_3 s_4 s_5 - s_2 s_3 c_4 c_5)^2 \cos^2\theta\;.
\end{align}
The scalars in 4-dimensions can be written in terms of the 3-dimensional scalar fields $\epsilon$ and $\varphi$ as
\begin{equation}
z_i=\;\epsilon_i - i\,e^{\varphi_i}\;,\;\;\;\;\;\;\;\;\;\;(i=\,1,2,3)
\end{equation}
and explicitly they read
\begin{align}
z_1=&\;\frac{2\,m\,\alpha\cos\theta\,(c_2s_3s_4c_5-s_2c_3c_4s_5)-\;i\,\rho^2}
        {\alpha^2\cos^2\theta+(r+2m{s_2}^2)(r+2m{s_5}^2)}\;,\nonumber\\
z_2=&\;z_1\,(2\leftrightarrow3)\;,\nonumber\\
z_3=&\;z_1\,(2\leftrightarrow4)\;.
\end{align}
To derive the $D=4$ metric and vector fields we use the dualization formulae (\ref{AF2})-(\ref{AF3}).
We can also \emph{locally} integrate $\mathbb{F}^M$ to a symplectic vector of electric and magnetic potentials $A^M_\mu$:
\begin{equation}
\mathbb{F}^M=dA^M\,\,;\,\,\,\,A^M=\mathcal{Z}^M\,(dt+\omega)+A^M_\varphi\,d\varphi\,,
\end{equation}
where $A^M_i\,dx^i=A^M_\varphi\,d\varphi$ are solutions to the differential equations (for the sake
 of notational simplicity we suppress the symplectic index $^M$):
\begin{align}
\partial_r A_\varphi=&\;-e_3\,e^{-2U}\,\mathbb{C}\,\mathcal{M}_{(4)}\,\partial^\theta\mathcal{Z}- \mathcal{Z}\,\partial_r \omega_\varphi\,,\nonumber\\
\partial_\theta A_\varphi=&\;e_3\,e^{-2U}\,\mathbb{C}\,\mathcal{M}_{(4)}\,\partial^r\mathcal{Z}- \mathcal{Z}\,\partial_\theta\omega_\varphi\,,
\end{align}
which directly follow from (\ref{AF2}) (indices are raised and lowered using $g^{(3)}$).
\newline
We find, for the $D=4$ metric (\ref{ds2}), the relevant quantities:
\begin{align}
e^{2U}=&\;\frac{\tilde\Delta}{\rho^2}\,\,\,;\,\,\,\,\,\,
\omega=\;\frac{2\,m\,\alpha\sin^2\theta\left(\,(P_c-P_s)\,r+2mP_s\,\right)}
         {\tilde\Delta}\,,
\end{align}
\newline
with the following expressions for the $3D$ scalars $Z^M$ fields
\begin{align}
Z_1=&\;\frac{1}{\rho^4}\left(\sqrt{2}\, m \alpha \cos\theta\,
   \left(c_2 c_3 c_4 s_5 (\alpha^2 \cos^2\theta + r (r+2 m {s_5}^2)) - s_2 s_3 s_4 c_5 (\alpha^2 \cos^2\theta +(r-2m)(r+2 m {s_5}^2))\right)\right)\,,\nonumber\\
Z_2=&\; \frac{1}{\rho^4}\,(\,2\sqrt{2}\, m^2 \alpha^2 \cos^2\theta\,((c_2^2+s_2^2)s_3c_3s_4c_4s_5c_5-s_2c_2(2s_3^2s_4^2s_5^2+s_3^2s_4^2+s_3^2s_5^2+s_4^2s_5^2))\,+\nonumber\\
   &+\sqrt{2}ms_2c_2(r\alpha^2\cos^2\theta+(r+2ms_3^2)(r+2ms_4^2)(r+2ms_5^2)))\;,\nonumber\\
\nonumber\\
Z_3=&\;Z_2\,(2\leftrightarrow3)\;,\;\;\;\;\;\;\;\;\;\;\;\;
Z_4=\;Z_2\,(2\leftrightarrow4)\;,\;\;\;\;\;\;\;\;\;\;\;\;
Z_5=\;Z_2\,(2\leftrightarrow5)\;,\nonumber\\
Z_6=&\;-\,Z_1\,(5\leftrightarrow2)\;,\;\;\;\;\;\;\;
Z_7=\;-\,Z_1\,(5\leftrightarrow3)\;,\;\;\;\;\;\;\;\;\,
Z_8=\;-\,Z_1\,(5\leftrightarrow4)\;.
\end{align}
\newline
The integration of the above equations yields the following result for the $\varphi$-components of the 4-dimensional vector fields:
\begin{align}
A^1_\varphi=&\;-\,\frac{\sqrt{2}\,m\,\Delta\,\cos\theta\,c_5s_5}{\tilde{\Delta}}\;,\nonumber\\
A^2_\varphi=&\;-\,\frac{\sqrt{2}\,m\,\alpha\,\sin^2\theta\,(c_2s_3s_4s_5(2m-r)+r\,s_2c_3c_4c_5)}{\tilde{\Delta}}\;,\nonumber\\
A^3_\varphi=&\;A^2_\varphi\,(2\leftrightarrow3)\;,\;\;\;\;\;\;\;\;\;\;\;\;
A^4_\varphi=\;A^2_\varphi\,(2\leftrightarrow4)\;,\;\;\;\;\;\;\;\;\;\;\;\;
A^5_\varphi=\;A^2_\varphi\,(2\leftrightarrow5)\;,\nonumber\\
A^6_\varphi=&\;-A^1_\varphi\,(5\leftrightarrow2)\;,\;\;\;\;\;\;\;\,
A^7_\varphi=\;-A^1_\varphi\,(5\leftrightarrow3)\;,\;\;\;\;\;\;\;\;\;
A^8_\varphi=\;-A^1_\varphi\,(5\leftrightarrow4)\;.
\end{align}
\newline
In this case, we find, from eq.s (\ref{allthat}), the following expression for the $ADM$-mass, the electric-magnetic charges $\Gamma^M=\;(p_0,\overrightarrow{p_i},q_0,\overrightarrow{q_i})$, the angular momentum $M_\varphi$ and the entropy $S$ (given by eq. (\ref{entropy}))
\begin{align}
M_{ADM}=&\;\frac{1}{8}\,m\,\left(\frac{1}{\beta_2}+\beta_2+\frac{1}{\beta_3}+\beta_3+\frac{1}{\beta_4}+\beta_4+\frac{1}{\beta_5}+\beta_5\right)\,,\nonumber\\
\Gamma^M=&\;\left(\frac{m(-1+{\beta_5}^2)}{2\sqrt{2}\,\beta_5},\,
       0,\,0,\,0,\,0,\,
       \frac{m(1-{\beta_2}^2)}{2\sqrt{2}\,\beta_2},\,
       \frac{m(1-{\beta_3}^2)}{2\sqrt{2}\,\beta_3},\,
       \frac{m(1-{\beta_4}^2)}{2\sqrt{2}\,\beta_4}\right)\;,\nonumber\\
M_\varphi=&\;m\,\alpha\,(P_c-P_s)\;,\nonumber\\
S=&\;2\pi\,m \left[m \,(P_c+P_s)+ c\, (P_c-P_s)\right]\,,
\end{align}
where $c=\sqrt{m^2-\alpha^2}$.
\newline

\subsection{Extremal Limits}
Let us start redefining:
\begin{equation}
\beta_{\ell'}\rightarrow m^{\sigma_{\ell'}}\beta_{\ell'}\;\;,\;\;\;\;\alpha\rightarrow m\Omega\;\;\;\;\;\;\;(\ell'=2,3,4,5)\\
\end{equation}
where $\sigma_{\ell'}=\;\pm1$, and introduce also the symbol $\zeta_\sigma=\prod_{\ell'}\sigma_{\ell'}$. Next, send $m$ to zero keeping the other parameters fixed.\\
There are 16 different ways to rescale the four $\beta_{\ell'}$-parameters and the general results for the extremal limits of the $ADM$-mass, electric-magnetic charges $\Gamma^M$, angular momentum $M_\varphi$, entropy $S$ and the quartic invariant $I_4$ are
\begin{align}
{M_{ADM}}^{(extr)}=&\;\frac{1}{8}\;\;\sum_{\ell'}
      \frac{1}{{\beta_{\ell'}}^{\sigma_{\ell'}}}\,,\nonumber\\
{\Gamma^M}^{(extr)}=&\;\left(\frac{-\,\sigma_5}{2\sqrt{2}\,{\beta_5}^{\sigma_5}},\,
       0,\,0,\,0,\,0,\,
       \frac{\sigma_2}{2\sqrt{2}\,{\beta_2}^{\sigma_2}},\,
       \frac{\sigma_3}{2\sqrt{2}\,{\beta_3}^{\sigma_3}},\,
       \frac{\sigma_4}{2\sqrt{2}\,{\beta_4}^{\sigma_4}}\right)\;,\nonumber\\
I_4=&\;-\,4\;{p_0}^e\;{q_1}^e\;{q_2}^e\;{q_3}^e\;=\;\frac{1}{16}\;\;\zeta_\sigma\;\prod_{\ell'}\frac{1}{{\beta_{\ell'}}^{\sigma_{\ell'}}}\;,\nonumber\\
{M_\varphi}^{(extr)}=&\;\frac{\Omega}{16}\;\;\sqrt{\prod_{\ell'}\frac{1}{{\beta_{\ell'}}^{\sigma_{\ell'}}}}\;\;(1-\zeta_\sigma)\;=
\;\frac{\Omega}{4}\;\;\sqrt{|I_4|}\;\;(1-\zeta_\sigma)\;,\nonumber\\
S^{(extr)}=&\;\pi\,\sqrt{|I_4|-4 ( M_{\varphi}^{(extr)})^2}=\pi\,\sqrt{|I_4|}\sqrt{1-\frac{1}{2}\Omega^2(1-\zeta_\sigma)}\,.
\end{align}
\newline
where we have used the short notation ${\Gamma^M}^{(extr)}=({p_0}^e,\,0,\,0,\,0,\,0,\,{q_1}^e,\,{q_2}^e,\,{q_3}^e)$ for the extremal charges. The solutions can be classified as
\begin{align}
  \textbf{BPS} &:\; I_4(p^e,q^e)>0\;\;\;\;\sigma_2=\sigma_3=\sigma_4\,;\nonumber\\
  \textbf{non-BPS 1} &:\; I_4(p^e,q^e)>0\;\;\;\;\sigma_2,\sigma_3,\sigma_4\;\; \mbox{not all equal}\,;\nonumber\\
  \textbf{non-BPS 2} &:\;I_4 (p^e,q^e)<0\,.\nonumber
\end{align}
\newline
Then, we can write the extremal limits for the 4-D vector fields ${A^M_\varphi}^{(extr)}$ as
\begin{align}
{A^{\ell'}_\varphi}^{(extr)}=&\;-\;\frac{\sqrt{2}}{r}\;{M^{(extr)}_\varphi}\,\sin^2\theta\;\;\zeta_\sigma\;\sigma_{\ell'}\;\;\;\;\;\;\;({\ell'}=2,3,4,5)\;,\nonumber\\
{A^\ell_\varphi}^{(extr)}=&\;-{\Gamma^\ell}^{(extr)}\cos\theta\;\;\;\;\;\;\;\;\;\;\;\;\;\;\;\;\;\;\;\;\;\;\;\;\;\;\;(\ell=1,6,7,8)\;.
\end{align}
\newline
We obtain, for the quantities involved in the metric expression, the following extremal limits
\begin{align}
{e^{2U}}^{(extr)}=&\;\left(H_2H_3H_4H_5-\frac{4\,\cos^2\theta({M^{(extr)}_\varphi})^2}{r^4}\right)^{-1/2}\,,\nonumber\\
\omega=&\;\frac{2\sin^2\theta\,{M^{(extr)}_\varphi}}{r}\;,
\end{align}
where, in the previous expressions, we have used the harmonic functions
\begin{align}
H_5=&\;1+\frac{\sqrt{2}\,|{p_0}^e|}{r}\;,\;\;\;\;\;
H_2=\;1+\frac{\sqrt{2}\,|{q_1}^e|}{r}\;,\nonumber\\
H_3=&\;1+\frac{\sqrt{2}\,|{q_2}^e|}{r}\;,\;\;\;\;\;
H_4=\;1+\frac{\sqrt{2}\,|{q_3}^e|}{r}\;.
\end{align}
\newline
The extremal limits for the 4-D $z$ scalar fields are
\begin{align}
z_1^{(extr)}=&\;-\sigma_2\sigma_5\;\frac{2\,\cos\theta\,{M^{(extr)}_\varphi}}{H_2H_5\,r^2}\;
       -\;i\;\frac{{e^{-2U}}^{(extr)}}{H_2H_5}\;,\nonumber\\
z_2^{(extr)}=&\;z_1\,(2\leftrightarrow3)\;,\nonumber\\
z_3^{(extr)}=&\;z_1\,(2\leftrightarrow4)\;,
\end{align}
while the limits for the 3-D scalar fields ${Z^M}^{(extr)}$ read
\begin{align}
{Z^1}^{(extr)}=&\;-\;\sigma_5\;\frac{\sqrt{2}\;H_5\;r^2\,\cos\theta\;{M^{(extr)}_\varphi}}
                          {H_2H_3H_4H_5\;r^4-4\,\cos^2\theta\;({M^{(extr)}_\varphi})^2}\,,\nonumber\\
{Z^2}^{(extr)}=&\;\;\sigma_2\;\frac{\;-H_3H_4H_5\;r^3\;|{q_1}^e|+2\sqrt{2}\;\cos\theta\;({M^{(extr)}_\varphi})^2}
                          {H_2H_3H_4H_5\;r^4-4\,\cos^2\theta\;({M^{(extr)}_\varphi})^2}\,,\nonumber\\
\\
{Z^3}^{(extr)}=&\;\;{Z^2}^{(extr)}\,(2\leftrightarrow3\,,\;\;|{q_1}^e|\rightarrow|{q_2}^e|)\,,\nonumber\\
{Z^4}^{(extr)}=&\;\;{Z^2}^{(extr)}\,(2\leftrightarrow4\,,\;\;|{q_1}^e|\rightarrow|{q_3}^e|)\,,\nonumber\\
{Z^5}^{(extr)}=&\;\;{Z^2}^{(extr)}\,(2\leftrightarrow5\,,\;\;|{q_1}^e|\rightarrow|{p_0}^e|)\,,\nonumber\\
{Z^6}^{(extr)}=&\;-\;{Z^1}^{(extr)}\,(5\leftrightarrow2)\,,\nonumber\\
{Z^7}^{(extr)}=&\;-\;{Z^1}^{(extr)}\,(5\leftrightarrow3)\,,\nonumber\\
{Z^8}^{(extr)}=&\;-\;{Z^1}^{(extr)}\,(5\leftrightarrow4)\;,
\end{align}

\section{The $q_0,\,p^i$\;-\;case}\label{q0pi}
Now let us consider the solution generated using the second subgroup of commuting generators
 $\mathcal{J}_\ell =\{\mathcal{J}_1\,,\mathcal{J}_6\,,\mathcal{J}_7\,,\mathcal{J}_8\}$,
that give an Harrison transformation of the form\footnote{For the $D=3$ description of this solution see \cite{Cvetic:2013cja}. }
\begin{equation}
\mathcal{O}_{(q_0,\,p^i)}=\;e^{\log(\beta_\ell)\;\mathcal{J}_\ell}\;,
\;\;\;\;\;\;\;\;\;\;\;\;\;\ell=1,6,7,8\;.
\end{equation}
to the Kerr solution, and solving the corresponding matrix equation
\begin{equation}
\mathcal{M}[\Phi^I(r,\theta)]=\mathcal{O}_{(q_0,\,p^i)}\;\mathcal{M}[\Phi^I_{(K)}(r,\theta)]\;\mathcal{O}_{(q_0,\,p^i)}^T\,.\label{mate
q0pi}
\end{equation}
Introduce now the combination of the $\beta$-parameters:
\begin{align}
c_\ell=&\;\frac{1+\beta_\ell}{2\,\sqrt{\beta_\ell}}\;,\;\;\;\;\;\;\;
s_\ell=\frac{-1+\beta_\ell}{2\,\sqrt{\beta_\ell}}\;\;\;\;\;\;\;\;\;\;\;(\ell=1,6,7,8)\nonumber\\
P_c=&\;c_1\,c_6\,c_7\,c_8\;,\;\;\;\; P_s=\;s_1\,s_6\,s_7\,s_8\;.\nonumber
\end{align}
and also the quantity
\begin{align}
\rho^4=&\;\left(\alpha^2 \cos^2\theta+(r + 2 m {s_1}^2) (r + 2 m{s_6}^2)\right)\,
        \left(\alpha^2 \cos^2\theta +(r + 2 m {s_7}^2) (r + 2 m{s_8}^2)\right)+\\
        &+4\alpha^2 m^2 (s_1 s_6 c_7 c_8 + c_1 c_6 s_7 s_8)^2 \cos^2\theta\;.
\end{align}
\newline
The 4-dimensional scalars $z_i = \epsilon_i - i\,e^{\varphi_i}$ can be now written as
\begin{align}
z_1=&\;\frac{\,-\,2\,m\,\alpha\cos\theta\,(s_1s_6c_7c_8+c_1c_6s_7s_8)-i\,\rho^2}
             {\alpha^2\cos^2\theta+(r+2m{s_7}^2)(r+2m{s_8}^2)}\;,\nonumber\\
z_2=&\;z_1\,(6\leftrightarrow 7)\;,\nonumber\\
z_3=&\;z_1\,(6\leftrightarrow 8)\;.\label{scal}
\end{align}
\newline
The relevant quantities for the $D=4$ metric reads:
\begin{align}
e^{2U}=&\;\frac{\tilde\Delta}{\rho^2}\,,\nonumber\\
\omega=&\frac{2\,m\,\alpha\sin^2\theta\,((P_c+P_s)\,r-2mP_s)}{\tilde\Delta}\,,
\end{align}
with the following expressions for the $3D$ scalars $Z^M$ fields
\begin{align}
Z_1=&\; -\frac{1}{\rho^4}\,(\,2\sqrt{2}\, m^2 \alpha^2 \cos^2\theta\,
   ((c_1^2+s_1^2) s_6 c_6 s_7 c_7 s_8 c_8 +
   s_1 c_1(2 s_6^2 s_7^2 s_8^2 +s_6^2 s_7^2 + s_7^2 s_8^2+ s_6^2 s_8^2))\,-\nonumber\\
   &+ \sqrt{2} m s_1 c_1 (r \alpha^2 \cos^2\theta +(r+2 m s_6^2)(r+2 m s_7^2)
   (r+2 m s_8^2))\,)\,,\nonumber\\
Z_2=&\; \frac{1}{\rho^4}(\,\sqrt{2}\, m \alpha \cos\theta\,
   (c_1 s_6 c_7 c_8 (\alpha^2 \cos^2\theta + r (r+2 m {s_6}^2)) +
   s_1 c_6 s_7 s_8 (\alpha^2 \cos^2\theta +(r-2m)(r+2 m {s_6}^2)))\,)\,,\nonumber\\
Z_3=&\;Z_2\,(6\leftrightarrow7)\;,\;\;\;\;\;\;\;
Z_4=\;Z_2\,(6\leftrightarrow8)\;,\;\;\;\;\;\;\;\;\;
Z_5=\;-\,Z_2\,(6\leftrightarrow1)\;,\nonumber\\
Z_6=&\;Z_1\,(1\leftrightarrow6)\;,\;\;\;\;\;\;\;
Z_7=\;Z_1\,(1\leftrightarrow7)\;,\;\;\;\;\;\;\;\;\,
Z_8=\;Z_1\,(1\leftrightarrow8)\;.
\end{align}
\newline
The $\varphi$-components of the 4-dimensional vector fields  are:
\begin{align}
A^1_\varphi=&\;\frac{\sqrt{2}\,m\,\alpha\,\sin^2\theta\,(c_1s_6s_7s_8(2m-r)-r\,s_1c_6c_7c_8)}{\tilde{\Delta}}\;,\nonumber\\
A^2_\varphi=&\;-\,\frac{\sqrt{2}\,m\,\Delta\,\cos\theta\,c_6s_6}{\tilde{\Delta}}\;,\nonumber\\
A^3_\varphi=&\;A^2_\varphi\,(6\leftrightarrow7)\;,\;\;\;\;\;\;\;\;
A^4_\varphi=\;A^2_\varphi\,(6\leftrightarrow8)\;,\;\;\;\;\;\;\;\;\;
A^5_\varphi=\;-A^2_\varphi\,(6\leftrightarrow1)\;,\nonumber\\
A^6_\varphi=&\;A^1_\varphi\,(1\leftrightarrow6)\;,\;\;\;\;\;\;\;\,
A^7_\varphi=\;A^1_\varphi\,(1\leftrightarrow7)\;,\;\;\;\;\;\;\;\;\;
A^8_\varphi=\;A^1_\varphi\,(1\leftrightarrow8)\;.\label{vect}
\end{align}
\newline
In this frame, we find the following expression for the $ADM$-mass, the electric-magnetic charges $\Gamma^M=\;(p_0,\overrightarrow{p_i},q_0,\overrightarrow{q_i})$, the angular momentum $M_\varphi$ and the entropy $S$ (given by eq. (\ref{entropy}))
\begin{align}
M_{ADM}=&\;\frac{1}{8}\,m\,(\frac{1}{\beta_1}+\beta_1+\frac{1}{\beta_6}+\beta_6+\frac{1}{\beta_7}+\beta_7+\frac{1}{\beta_8}+\beta_8)\,,\nonumber\\
\Gamma^M=&\;(\,0,\frac{m(-1+{\beta_6}^2)}{2\sqrt{2}\,\beta_6},\,
       \frac{m(-1+{\beta_7}^2)}{2\sqrt{2}\,\beta_7},\,
       \frac{m(-1+{\beta_8}^2)}{2\sqrt{2}\,\beta_8},\,
       \frac{m(1-{\beta_1}^2)}{2\sqrt{2}\,\beta_1},\,0,\,0,\,0\,)\;,\nonumber\\
M_\varphi=&\;m\,\alpha\,(P_c+P_s)\;,\nonumber\\
S=&\;2\pi\,m \left[m \,(P_c-P_s)+ c\, (P_c+P_s)\right]\,,
\end{align}
where $c=\sqrt{m^2-\alpha^2}$.
\newline

\subsection{Extremal Limits}\label{eliq0pi}
Let us redefine:
\begin{equation}
\beta_\ell\rightarrow\;m^{\sigma_\ell}\beta_\ell\;\;,\;\;\;\;\alpha\rightarrow\,m\Omega\;\;\;\;\;\;\;
(\ell=1,6,7,8)\\
\end{equation}
where $\sigma_\ell=\;\pm1$ and $\zeta_\sigma=\prod_{\ell}\sigma_\ell$. Then, send $m$ to zero keeping the other parameters fixed.\\
We find again 16 different ways to rescale the four $\beta_\ell$-parameters and the results for the extremal limits of the $ADM$-mass, electric-magnetic charges $\Gamma^M$, angular momentum $M_\varphi$, entropy $S$ and the quartic invariant $I_4$ read
\begin{align}
{M_{ADM}}^{(extr)}=&\;\frac{1}{8}\sum_\ell
      \frac{1}{{\beta_\ell}^{\sigma_\ell}}\,,\nonumber\\
{\Gamma^M}^{(extr)}=&\;(\,0,\frac{-\sigma_6}{2\sqrt{2}\,{\beta_6}^{\sigma_6}},\,
       \frac{-\sigma_7}{2\sqrt{2}\,{\beta_7}^{\sigma_7}},\,
       \frac{-\sigma_8}{2\sqrt{2}\,{\beta_8}^{\sigma_8}},\,
       \frac{\sigma_1}{2\sqrt{2}\,{\beta_1}^{\sigma_1}},\,0,\,0,\,0)\;,\nonumber\\
I_4=&\;4\;{q_0}^e\;{p_1}^e\;{p_2}^e\;{p_3}^e\;=\;-\;\frac{1}{16}\;\zeta_\sigma\;\prod_\ell
             \frac{1}{{\beta_\ell}^{\sigma_\ell}}\;,\nonumber\\
{M_\varphi}^{(extr)}=&\;\frac{\Omega}{16}\,\sqrt{\prod_\ell\frac{1}{{\beta_\ell}^{\sigma_\ell}}}\;\;(1+\zeta_\sigma)\;=
\;\frac{\Omega}{4}\,\sqrt{|I_4|}\;\;(1+\zeta_\sigma)\;,\nonumber\\
S^{(extr)}=&\;\pi\,\sqrt{|I_4|-4 ( M_{\varphi}^{(extr)})^2}=\pi\,\sqrt{|I_4|}\sqrt{1-\frac{1}{2}\Omega^2(1+\zeta_\sigma)}\,.
\end{align}
\newline
where we have used now the short notation ${\Gamma^M}^{(extr)}=(0,\,{p_1}^e,\,{p_2}^e,\,{p_3}^e,\,{q_0}^e,\,0,\,0,\,0)$ for the extremal charges. Also in this case the solutions can be classified as
\begin{align}
  \textbf{BPS}
     &:\;I_4(p^e,q^e)>0\;\;\;\;\sigma_6=\sigma_7=\sigma_8\,;\nonumber\\
  \textbf{non-BPS 1} &:\; I_4(p^e,q^e)>0\;\;\;\;\sigma_6,\sigma_7,\sigma_8\;\;
     \mbox{not all equal}\,;\nonumber\\
  \textbf{non-BPS 2} &:\;I_4 (p^e,q^e)<0\,.\nonumber
\end{align}
\newline
The extremal limits for the 4-D vector fields ${A^M_\varphi}^{(extr)}$ can be written
\begin{align}
{A^\ell_\varphi}^{(extr)}=&\;\frac{\sqrt{2}}{r}\;{M^{(extr)}_\varphi}\,\sin^2\theta\;\;\zeta_\sigma\;\sigma_\ell\;\;\;\;\;\;\;(\ell=1,6,7,8)\;,\nonumber\\
{A^{\ell'}_\varphi}^{(extr)}=&\;-{\Gamma^{\ell'}}^{(extr)}\cos\theta\;\;\;\;\;\;\;\;\;\;\;\;\;\;\;\;\;\;\;\;(\ell'=2,3,4,5)\;.
\end{align}
\newline
We obtain, for the quantities involved in the metric expression, the following extremal limits
\begin{align}
{e^{2U}}^{(extr)}=&\;\left(H_1H_6H_7H_8-\frac{4\,\cos^2\theta({M^{(extr)}_\varphi})^2}{r^4}\right)^{-1/2}\,,\nonumber\\
\omega=&\;\;\frac{2\sin^2\theta\,{M^{(extr)}_\varphi}}{r}
\end{align}
where, in the previous expressions, we have used the harmonic functions
\begin{align}
H_6=&\;1+\frac{\sqrt{2}\,|{p_1}^e|}{r}\;,\;\;\;\;\;
H_7=\;1+\frac{\sqrt{2}\,|{p_2}^e|}{r}\;,\nonumber\\
H_8=&\;1+\frac{\sqrt{2}\,|{p_3}^e|}{r}\;,\;\;\;\;\;
H_1=\;1+\frac{\sqrt{2}\,|{q_0}^e|}{r}\;.
\end{align}
\newline
The limits for the 4-D $z$ scalar fields read
\begin{align}
z_1^{(extr)}=&\;-\sigma_1\sigma_6\;\frac{2\,\cos\theta\,{M^{(extr)}_\varphi}}{H_7H_8\,r^2}\;
       -\;i\;\frac{{e^{-2U}}^{(extr)}}{H_7H_8}\;,\nonumber\\
z_2^{(extr)}=&\;z_1\,(6\leftrightarrow7)\;,\nonumber\\
z_3^{(extr)}=&\;z_1\,(6\leftrightarrow8)\;,
\end{align}
\newline
while the limits for the 3-D scalar fields ${Z^M}^{(extr)}$ are
\begin{align}
{Z^1}^{(extr)}=&\;-\;\sigma_1\;\frac{\;H_6H_7H_8\;r^3\;|{q_0}^e|+2\sqrt{2}\;\cos\theta\;\left({M^{(extr)}_\varphi}\right)^2}
                          {H_1H_6H_7H_8\;r^4-4\,\cos^2\theta\;\left({M^{(extr)}_\varphi}\right)^2}\,,\nonumber\\
{Z^2}^{(extr)}=&\;-\;\sigma_6\;\frac{\sqrt{2}\;H_6\;r^2\,\cos\theta\;{M^{(extr)}_\varphi}}
                          {H_1H_6H_7H_8\;r^4 -4\,\cos^2\theta\;\left({M^{(extr)}_\varphi}\right)^2}\,,\nonumber\\
\\
{Z^3}^{(extr)}=&\;\;{Z^2}^{(extr)}\,(6\leftrightarrow7)\,,\nonumber\\
{Z^4}^{(extr)}=&\;\;{Z^2}^{(extr)}\,(6\leftrightarrow8)\,,\nonumber\\
{Z^5}^{(extr)}=&\;-\;{Z^2}^{(extr)}\,(1\leftrightarrow6)\,,\nonumber\\
{Z^6}^{(extr)}=&\;\;{Z^1}^{(extr)}\,(1\leftrightarrow6\,,\;\;|{q_0}^e|\rightarrow|{p_1}^e|)\,,\nonumber\\
{Z^7}^{(extr)}=&\;\;{Z^1}^{(extr)}\,(1\leftrightarrow7\,,\;\;|{q_0}^e|\rightarrow|{p_2}^e|)\,,\nonumber\\
{Z^8}^{(extr)}=&\;\;{Z^1}^{(extr)}\,(1\leftrightarrow8\,,\;\;|{q_0}^e|\rightarrow|{p_3}^e|)\;,
\end{align}

\section{The $5^{th}$ Parameter and Concluding Remarks} Although the main focus of this note is the geometrical relationship between
$G$-orbits of black holes, we observe that the extremal static and under-rotating solutions found above are 4-parameter solutions,
the parameters being related to the four charges in the two normal forms $\{q_0,\,p^i\}$ and $\{p^0,\,q_i\}$. It is known,
see for instance  \cite{attractors1,Andrianopoli:1997wi,Gimon:2007mh}, that the most general solution to the symmetric
supergravities considered here (which have the STU model as a consistent truncation),
modulo action of $G_4$ (i.e. the \emph{seed solution} with respect to the action of $G_4$), has 5 independent parameters. These can be written in terms of
 five independent $H_4$-invariants computed at radial infinity (depending on $\phi_0^s$ and $p^\Lambda,\,q_\Lambda$).\footnote{In other words, in these models, one can define a maximal set of five functionally independent functions  $\mathcal{I}_1,\dots,\mathcal{I}_5$ of $\phi_0^s$ and $\Gamma^M$ which are invariant under the action of $G_4$ on both the scalar fields at infinity and the electric-magnetic charges.} This number 5 is nothing but the rank $p$ of $H^*/H_c$, introduced in Section \ref{el}, plus one
 (in the $T^3$-model
 $p=2$ and the seed solution with respect to $G_4$ is a three-parameter one). In the $D=3$
 description, a larger symmetry  group $G$ is manifest. In particular on the charges we can act by means of the group
 $H_c={\rm U}(1)_E\times H_4$ which contains, besides $H_4$, an additional ${\rm U}(1)_E$-symmetry. Using it we can reduce the number
  of independent invariants characterizing the solution from $p+1=5$ to $p=4$, so that the \emph{seed solution} with respect to $G$
  is a \emph{four-parameter solution} characterized by electric and magnetic charges in one of the two  normal forms of $\Gamma^M$
   with respect to $H_c$: $\{q_0,\,p^i\}$ and $\{p^0,\,q_i\}$. In support of this argument we observe that the nilpotent
   $H^*$-orbits of $Q$, corresponding to the extremal, regular, single-center solutions are unique and contain
   the 4-parameter solutions constructed here (see for instance \cite{nilorbits}).\par
 We have explicitly checked in the STU model that, acting on the 4-parameter BPS and non-BPS extremal solutions by means of a
 combination of ${\rm U}(1)_E$
 and Harrison transformations, the $5^{th}$ parameter can be generated.
 In the STU model the 5 $H_4$-invariants can be constructed out of the central  and matter charges ($Z(\phi^s,p,q)$,
 $Z_{\bf I}(\phi^s,p,q)$, ${\bf I}=1,2,3$), in terms of their moduli and overall phase and read (in the chosen symplectic frame):
 \begin{equation}
\mbox{5 invariants}\,\,\,=\,\,\,\,\, \{|Z|,\,|Z_{\bf I}|,\,{\rm Arg}(Z\bar{Z}_1\bar{Z}_2\bar{Z}_3)\}\,.\label{invars}
 \end{equation}
where the  central and matter charges are defined as (see Appendix \ref{stumodel}):
 \begin{equation}
 Z=-V^T\mathbb{C}\,\Gamma\,\,\,;\,\,\,\,Z_{\bf I}=-e_{\bf I}{}^i\,D_iV^T\mathbb{C}\,\Gamma\,,\label{ccs}
 \end{equation}
 $e_{\bf I}{}^i$ being the inverse complex vielbein on $\mathcal{M}^{(4)}_{scal}$ and $V^M(\phi^s)$ is the
 covariantly holomorphic section of the symplectic bundle on the manifold.
 If we start from the solutions with charges in the normal form $q_0,\,p^i$, we apply to it an Ehlers transformation
 $\mathcal{O}_E(\alpha)$
 followed by a Harrison one $\mathcal{O}_H(v_{\ell'})$ in $\exp(\mathbb{J}^{(N)})$, where $\mathbb{J}^{(N)}$
 corresponds to the
  other normal form, namely $p^0,\, q_i$. The Harrison parameters are then determined in terms of $\alpha$,
   $v_{\ell'}=v_{\ell'}(\alpha)$,
   in order to cancel the NUT charge produced by $\mathcal{O}_E(\alpha)$.
  The resulting transformation generating the $5^{th}$-parameter reads then:
 \begin{equation}
 \mathcal{O}_{5^{th}\,par.}=\mathcal{O}_E(\alpha)\,\mathcal{O}_H(v_{\ell'}(\alpha))\,.
 \end{equation} We have checked on the extremal solutions that
   the five invariants (\ref{invars}) are independent functions of $q_0,p^i$ and $\alpha$ and therefore conclude that  \emph{ $5^{th}$ parameter can be generated by means of $G$}. We refer the reader to Appendix \ref{5par} for an explicit calculation.\par
   As a general comment, let us observe that in order to find the 5-parameter solution we had to perform a set of non-commuting Ehlers and Harrison transformations on the 4-parameter solution, \emph{whose net effect is to modify   topological properties of the $D=4$ black hole}. More precisely, we introduced the 5th parameter by a NUT-charge-generating $U(1)_E$ transformation in $D=3$, and then converted it, by an appropriate Harrison transformation into a gauge charge non-commuting with the other gauge charges. In the $D=3$ description, where all the bosonic degrees of freedom  of the stationary black hole solution (corresponding to the metric, the gauge vectors and the $D=4$ scalars) are collectively described by the scalar sigma-model $G/H^*$, the above prescription is among  the  allowed symmetry transformations on the set of conserved charges. However, in the $D=4$ description this  transformation is highly non trivial: It generates the 5th parameter as a NUT charge, that is as a non-trivial topology of space-time, and then  (in order to have an asymptotically flat black hole solution) trades it  into a gauge charge thus adding to the non triviality of  the gauge bundle. In our setting we have chosen to fix the scalars at radial infinity to their origin, otherwise, for the extremal $I_4<0$ black hole, the same solution could have been converted, by the action of $G_4/H_4$, into one where the gauge bundle has commuting charges but the axions acquire a non trivial value at radial infinity \cite{Lopes Cardoso:2007ky,Hotta:2007wz,Gimon:2007mh,Goldstein:2008fq}. Instead of referring to the $D=3$ description, the other way adopted in \cite{Gaiotto:2005gf,Lopes Cardoso:2007ky,Goldstein:2008fq},  to find the $D=4$ seed solution has been via Kaluza--Klein reduction from $D=5$. Also in this case, the seed solution of $D=4$ stationary, asymptotically flat black holes was found to correspond to a $5D$ NUT-charge configuration with angular momentum.

We conclude that in all its descriptions the seed, 5-parameter, solution should have an additional non-trivial topological feature with respect to the 4-parameter one.
This distinction, at least for the $I_4<0$ extremal black hole in the static case, reflects itself in the different behavior of the harmonic functions ${ \bf H}=(H^M)$ characterizing the solution:

In the 4-parameter solution they obey the relation ${\bf H}^T \cdot \mathbb{C} \cdot \partial_r \mathbf{H}=0$, while the  5-parameter seed solution satisfies $\mathbf{H}^T \cdot \mathbb{C} \cdot \partial_r \mathbf{H}\neq 0$. This shows that the transformation connecting the two cannot be a $D=4$ global  symmetry which would leave the symplectic product unaltered.

\vskip 5mm
   \par

The study of the extremal limits, started in \cite{Andrianopoli:2013kya} and concluded here, was also a
testing ground for the newly defined $\mathfrak{g}$-valued matrix $Q_\psi$, which encodes the rotation property
of the solution and which allows to directly compute the action of the symmetry group $G$ on the angular momentum $M_\varphi $.
We have seen that, in spite of having the manifestly $G_4$-invariant expression  in (\ref{MI4}), this quantity
 is far from being $G$-invariant. In the non-extremal case even  the manifest $G_4$-invariance of both $M_\varphi$ and of the entropy $S$, as functions of the electric-magnetic charges alone,
  is lost. We have argued at the end of Sect. \ref{el} that, if we retain the dependence of these two quantities from the boundary
  values $\phi_0^s$ of the scalar fields at radial infinity (that we have fixed to zero in the present analysis),
   then as functions of both $\phi_0^s$ and $\Gamma^M$,
 they could still be $G_4$-invariant. This was proven for $M_\varphi$ at the end of Sect. \ref{kf} on general grounds.
 As pointed out earlier, in the class of models we have been considering here there are five independent $G_4$-invariant functions $(\mathcal{I}_n)=(\mathcal{I}_1,\dots,\mathcal{I}_5)$ of $\phi_0^s$ and $\Gamma^M$, which reduce to those in (\ref{invars}) once we restrict to the STU truncation.
 We leave  the determination of the explicit expression of $M_\varphi,\,M_{ADM}$ and $S$ in terms of $\mathcal{I}_n$ in the Kerr-orbit to a future investigation. We just notice here that, once we solve this problem for the STU model, the same expressions in terms of $\mathcal{I}_n$ hold for all the other  symmetric models.
\section{Acknowledgements}
We wish to thank R. D'Auria for enlightening discussions.  This work was partially supported by
the Italian MIUR-PRIN contract 2009KHZKRX-007 "Symmetries of the Universe and of
the Fundamental Interactions".

\appendix

\section{The STU model}\label{stumodel}
The STU model is an $\mathcal{N}=2$ supergravity coupled to three vector multiplets ($n_s=6,\,n_v=4$) and with:
\begin{equation}
\mathcal{M}^{(D=4)}_{scal}=\frac{G_4}{H_4}=\left(\frac{{\rm SL}(2,\mathbb{R})}{{\rm SO}(2)}\right)^3\,.
\end{equation}
This manifold is a complex spacial K\"ahler space spanned by three complex scalar fields $z^a=\{S,T,U\}$.
The $D=4$ scalar metric for the STU model reads
\begin{equation}
dS^2_4=g_{rs}\,d\phi^sd\phi^r=2\,g_{a\bar{b}}dz^ad\bar{z}^{\bar{b}}=-2\,\sum_{a=1}^3\frac{dz^{a}
d\bar{z}^{\bar{a}}}{(z^a-\bar{z}^{\bar{a}})^2}=\sum_{{\bf I}=1}^3 e_i{}^{\bf I}\bar{e}_{\bar{\imath}}{}^{\bf I}\,dz^i\,d\bar{z}^{\bar{\imath}}\,.
\end{equation}
We also consider the real parametrization $\{\phi^s\}=\{\epsilon_i,\,\varphi_i\}$, related to the complex one by:
$z_i=\epsilon_i-i\,e^{\varphi_i}$. The K\"ahler potential has the simple form: $e^{-K}=8\,e^{\varphi_1+\varphi_2+\varphi_3}$. In the chosen symplectic frame (i.e. the special coordinate frame originating from Kaluza Klein reduction from $D=5$), the special geometry of $\mathcal{M}^{(D=4)}_{scal}$ is characterized by a holomorphic prepotential $\mathcal{F}(z)=z_1 z_2 z_3$. The holomorphic $\Omega^M(z)$ section of the symplectic bundle reads:
\begin{equation}
\Omega^M(z)=\{1,z_1,z_2,z_3,-z_1 z_2 z_3,z_2 z_3,z_1 z_3,z_1
   z_2\}\,,\label{omega}
\end{equation}
while the covariantly holomorphic section is given by $V^M(z,\bar{z})=e^{\frac{K}{2}}\,\Omega^M(z)$. In terms of
$V^M$ and of its covariant derivatives $D_i$ ($D_iV:=\partial_i V+\frac{\partial_i K}{2}\,V$) we write the central and matter charges (\ref{ccs}) of a black hole solution with quantized charges $\Gamma=(\Gamma^M)=(p^\Lambda,\,q_\Lambda)$:
 \begin{align}
 Z&=-V^T\mathbb{C}\,\Gamma=e^{\frac{K}{2}}\,(-q_0-q_1 z_1-q_2 z_2+p^3 z_1 z_2-q_3 z_3+p^2 z_1 z_3+p^1 z_2 z_3-p^0 z_1 z_2 z_3)\,,\nonumber\\
 Z_1&=-e_{ 1}{}^i\,D_iV^T\mathbb{C}\,\Gamma=-i\,e^{\frac{K}{2}}\,\left(q_0+q_2 z_2+q_3 z_3-p^1 z_2 z_3+q_1 \bar{z}_1-p^3 z_2 \bar{z}_1-p^2 z_3 \bar{z}_1+p^0 z_2 z_3 \bar{z}_1\right)\,,\nonumber\\
 Z_2&=-e_{ 2}{}^i\,D_iV^T\mathbb{C}\,\Gamma=-i\,e^{\frac{K}{2}}\,\left(q_0+q_1 z_1+q_3 z_3-p^2 z_1 z_3+q_2 \bar{z}_2-p^3 z_1 \bar{z}_2-p^1 z_3 \bar{z}_2+p^0 z_1 z_3 \bar{z}_2\right)\,,\nonumber\\
 Z_3&=-e_{ 3}{}^i\,D_iV^T\mathbb{C}\,\Gamma=-i\,e^{\frac{K}{2}}\,\left(q_0+q_1 z_1+q_2 z_2-p^3 z_1 z_2+q_3 \bar{z}_3-p^2 z_1 \bar{z}_3-p^1 z_2 \bar{z}_3+p^0 z_1 z_2 \bar{z}_3\right)\,.
 \end{align}
Let us also give the explicit form of the quartic invariant for the STU model:
\begin{align}
I_4(p,q)&=-(p^0)^2 q_0^2-2 \left(-2 p^1 p^2 p^3+p^0 q_3 p^3+p^0 p^1 q_1+p^0 p^2 q_2\right) q_0-(p^1)^2 q_1^2-\left(p^2 q_2-p^3 q_3\right)^2+\nonumber\\
&+2 q_1
   \left(p^1 p^3 q_3+q_2 \left(p^1 p^2-2 p^0 q_3\right)\right)\,.
\end{align}
Upon timelike reduction to $D=3$ the scalar manifold has the form $G/H^*$ with $G={\rm SO}(4,4)$ and $H^*={\rm SO}(2,2)^2$. We describe the generators of $\mathfrak{g}=\mathfrak{so}(4,4)$ in terms of Cartan $H_\alpha$ and shift generators $E_{\pm\alpha}$
in the fundamental representation, with the usual normalization convention:
\begin{equation}
[H_\alpha,\,E_{\pm \alpha}]=\pm 2\,E_{\pm \alpha}\,\,\,;\,\,\,\,[E_\alpha,\,E_{-\alpha}]=H_\alpha\,.
\end{equation}
In our notation $E_{-\alpha}=E_\alpha^\dagger=E_\alpha^T$. The positive roots of $\mathfrak{g}$ split into: the root $\beta_0$ of the Ehlers subalgebra $\mathfrak{sl}(2,\mathbb{R})_E$ commuting with the algebra $\mathfrak{g}_4$ of $G_4$ inside $\mathfrak{g}$; the roots $\alpha_i,\,(i=1,2,3)$ of $\mathfrak{g}_4$ and eight roots $\gamma_M$, $m=1,\dots,8$. The special coordinate parametrization of $\mathcal{M}^{(4)}_{scal}$ corresponds to a solvable parametrization of the manifold in which the real coordinates $(\phi^s)=(\epsilon_i,\,\varphi_i)$ are parameters of a solvable Lie algebra generated by $(T_s)=(E_{\alpha_i},\,\frac{1}{2}\,H_{\alpha_i})$. The coset representative $L_4$ is an element of the corresponding solvable group
\cite{solvable} defined by the following exponentialization prescription:
\begin{equation}
L_4(\phi^s)=\exp(\phi^s\,T_s)=\prod_{i=1}^3 e^{\epsilon_i E_{\alpha_i}}e^{\varphi_i \frac{H_{\alpha_i}}{2}}\,.\label{L4}
\end{equation}
The solvable (or Borel) subalgebra $Solv={\rm Span}(T_\mathcal{A})$, $\{T_\mathcal{A}\}=\{H_0,\,T_\bullet,\,T_s,\,T_M\}$ of $\mathfrak{g}$ used to define the parametrization of $\mathcal{M}_{scal}$ in terms of the $D=3$ scalars $\phi^I$ through the coset representative (\ref{cosetr3}), is defined by the identification:
\begin{equation}
H_0=\frac{H_{\beta_0}}{2}\,\,;\,\,\,\,T_\bullet=E_{\beta_0}\,\,\,;\,\,\,\,T_M=E_{\gamma_M}\,.
\end{equation}
The symplectic representation of $T_s$ in the duality representation ${\bf R}={\bf (2,2,2)}$ of $G_4$ is defined through their adjoint action on $T_M$: $[T_s,\,T_M]=-T_{s\,M}{}^N\,T_N$. In order to reproduce the form of the $T_{s\,M}{}^N$ in the chosen special coordinate frame (\ref{omega}), the generators $T_M$ corresponding to the roots $\gamma_M$, have to be ordered according to (\ref{gammaord}). In this basis, the symplectic representation of  $L_4=(L_{4\,M}{}^N)$ defined in (\ref{L4}) allows to define the matrix $\mathcal{M}_{(4)}$:
\begin{equation}
\mathcal{M}_{(4)\,MN}=-\sum_{P=1}^8(L_{4\,M}{}^P)(L_{4\,N}{}^P)\,.
\end{equation}
We give, for the sake of completeness, the matrix form of $\phi^s\,T_s$ in the symplectic representation ${\bf R}$:
\begin{align}
\phi^s\,T_s&=\sum_{i=1}^3\epsilon_i E_{\alpha_i}+\varphi_i \frac{H_{\alpha_i}}{2}=\left(\begin{matrix}A & B \cr {\bf 0} & -A^T\end{matrix}\right)\,,\nonumber\\
A&=\left(
\begin{array}{llll}
 \frac{\varphi _1}{2}+\frac{\varphi _2}{2}+\frac{\varphi _3}{2} & -\epsilon _1 & -\epsilon _2 & -\epsilon _3 \\
 0 & -\frac{\varphi _1}{2}+\frac{\varphi _2}{2}+\frac{\varphi _3}{2} & 0 & 0 \\
 0 & 0 & \frac{\varphi _1}{2}-\frac{\varphi _2}{2}+\frac{\varphi _3}{2} & 0 \\
 0 & 0 & 0 & \frac{\varphi _1}{2}+\frac{\varphi _2}{2}-\frac{\varphi _3}{2}
\end{array}
\right)\,.\nonumber\\
B&= \left(
\begin{array}{llll}
 0 & 0 & 0 & 0 \\
 0 & 0 & -\epsilon _3 & -\epsilon _2 \\
 0 & -\epsilon _3 & 0 & -\epsilon _1 \\
 0 & -\epsilon _2 & -\epsilon _1 & 0
\end{array}
\right)\,.
\end{align}
The pseudo-Cartan involution $\sigma$ defining the decomposition of $\mathfrak{g}$ into $\mathfrak{H}^*$ and $\mathfrak{K}^*$ is defined by the matrix $\eta=(-1)^{2 H_0}$.

\section{Generating the $5^{th}$ Parameter}\label{5par}
We give the details of the calculation only in the $I_4<0$ extremal solution.
We start from the under-rotating, non-BPS extremal solution having $I_4<0$, with charges in the normal form $q_0,\,p^i$. This was derived in Subsect. \ref{eliq0pi} taking, for instance,  $\sigma_\ell\equiv +1$. We then apply the transformation:
\begin{equation}
\mathcal{O}_{5^{th}\,param}=\exp(\alpha\,J_\bullet)\,\exp\left(\log(x(\alpha))\,\mathcal{J}_2+\log(x(\alpha))\,\mathcal{J}_3
+\log(x(\alpha))\,\mathcal{J}_4-\log(x(\alpha))\,\mathcal{J}_5\right)\,,
\end{equation}
where $J_\bullet:=E_{\beta_0}-E_{-\beta_0}$ is the generator of ${\rm U}(1)_E$ and $x(\alpha)$ is solution to the equation:
\begin{equation}
\sin(\alpha)=\frac{1-x^4}{x^4+6 x^2+1}\,.
\end{equation}
The above relation is derived by the condition of vanishing NUT charge.
Since for $x>0$ there is a one to one correspondence between $\sin(\alpha)$ and $x$, instead of computing $x$ as a function of $\alpha$ it is more convenient to express the latter in terms of the former and to substitute in $Q$ and $Q_\psi$. The resulting solution is still defined by vanishing  scalar fields at radial infinity $\phi_0^s=0\,\,\Leftrightarrow\,\,\,z_j=-i$, and its magnetic and  electric charges read:
\begin{align}
p_0=&\;-\frac{x^2-1}{2\sqrt{2}\;(x^4+6x^2+1)^2}\;\left((x^2-1)^2\frac{1}{\beta_1}\;+\;(3x^4+10x^2+3)\left(\frac{1}{\beta_6}+\frac{1}{\beta_7}+\frac{1}{\beta_8}\right)\right)\,,\nonumber\\
p_1=&\;-\frac{x\;\sqrt{3x^4+10x^2+3}}{2\sqrt{2}\;(x^4+6x^2+1)^2}\,\left((x^2-1)^2\left(\frac{1}{\beta_6}+\frac{1}{\beta_7}+\frac{1}{\beta_8}\right)\;+\;(3x^4+10x^2+3)\frac{1}{\beta_6}\right)\,,\nonumber\\
p_2=&\;p_1\,(6\leftrightarrow 7)\;,\nonumber\\
p_3=&\;p_1\,(6\leftrightarrow 8)\;,\nonumber\\
q_0=&\;-\,p_1\,(6\leftrightarrow1)\;,\nonumber\\
q_1=&\;p_0\,(1\leftrightarrow6)\;,\nonumber\\
q_2=&\;p_0\,(1\leftrightarrow7)\;,\nonumber\\
q_3=&\;p_0\,(1\leftrightarrow8)\;.
\end{align}
Computing the five invariants (\ref{invars}) with the above charges one can verify that they are
independent functions of $\beta_\ell,\,x$.
The ADM mass and  angular momentum read:
\begin{align}
M_{ADM}&=\frac{x \left(x^2+1\right)^3}{\left(x^4+6 x^2+1\right)^2} \left(\frac 1{\beta_1}+\frac 1{\beta_6}+
\frac 1{\beta_7}+
\frac 1{\beta_8}
\right) \,,\nonumber\\
M_\varphi&=\frac{x^2 \Omega }{\left(x^4+6 x^2+1\right) \sqrt{\beta_1} \sqrt{\beta_6} \sqrt{\beta_7}
   \sqrt{\beta_8}}=\frac{\Omega}{2}\,\sqrt{|I_4(p,q)|}\,,
\end{align}
thus confirming the general eq. (\ref{MI4}) on the five parameter solution.

\end{document}